\documentclass[american,amssymb,amsmath]{revtex4}
\usepackage[T1]{fontenc}
\usepackage[latin9]{inputenc}
\usepackage{amsmath,amsthm}
\usepackage{amssymb}
\usepackage{esint}
\usepackage[all, knot]{}
\makeatletter
\@ifundefined{textcolor}{}
{%
 \definecolor{BLACK}{gray}{0}
 \definecolor{WHITE}{gray}{1}
 \definecolor{RED}{rgb}{1,0,0}
 \definecolor{GREEN}{rgb}{0,1,0}
 \definecolor{BLUE}{rgb}{0,0,1}
 \definecolor{CYAN}{cmyk}{1,0,0,0}
 \definecolor{MAGENTA}{cmyk}{0,1,0,0}
 \definecolor{YELLOW}{cmyk}{0,0,1,0}
 }
\newtheorem{theorem}{Theorem}
\newtheorem{definition}{Definition}

\newtheorem{corollary}{Corollary}
\newtheorem{conjecture}{Conjecture}
\usepackage[all, knot]{xy}

\makeatother

\usepackage{amstext}
\usepackage{amssymb}

\begin{document}

\title{Magnetic monopoles, squashed 3-spheres and gravitational instantons from exotic $\mathbb{R}^4$}

\author{Torsten Asselmeyer-Maluga}

\email{torsten.asselmeyer-maluga@dlr.de}

\affiliation{German Aero Space Center, Rutherfordstr. 2, 12489 Berlin}

\author{Jerzy Kr\'ol}

\email{iriking@wp.pl}

\affiliation{University of Silesia, Institute of Physics, ul. Uniwesytecka 4,
40-007 Katowice}
\begin{abstract}
We show that, in some limit, gravitational instantons correspond to exotic smooth geometry on $\mathbb{R}^4$. The geometry of this exotic $\mathbb{R}^4$ represents magnetic monopoles of Polyakov-'t Hooft type and the BPS condition allows for the generation of the charges from the gravitational sources of exotic $\mathbb{R}^4$.  Higgs field is, as usual, present in the monopole configurations. We indicate some possible scenarios in cosmology, particle physics and condensed matter where pure $SU(2)$ Yang-Mills theory on the exotic $\mathbb{R}^4$ aquires non-zero mass for its gauge boson when the smoothness is changed to the standard $\mathbb{R}^4$ and, then, it is described by Yang-Mills Higgs theory. The derivation of the results is based on the quasi-modularity of the expressions and the geometry of foliations in the limit.  

\end{abstract}


\maketitle

\section{Introduction}
Foliations are quite remarkable objects in geometry and topology.
They refine the structure of fiber bundles and fibrations in some
cases. They are purely classical geometric 'tiling' of a manifold
by (lower-dimensional) leaves. On the other hand, from the point of
view of the geometry of their spaces of leaves, foliations are 'quantum'
objects and were recognized by Connes as the basic instances of the
non-commutative geometries. He assigned to every foliation a (convolution)
$C^{\star}$-algebra of operators which in many cases is non-commutative.
But when (quantum) field theory is formulated on a manifold $M$ then,
usually, one forgets the foliations and 'purely classical' geometric
structures on $M$ overwhelm our practice and thinking. This is rather
natural since physical results do not depend on foliations. Is it
true, indeed? In general relativity (GR) on Lorentzian manifolds the
global hyperbolic structure is nothing but the choice of certain 'foliation'
of 4-manifold via 3-dimensional slices (leaves). Many other results,
especially in GR, rely heavily on such foliations of spacetime (for
example ADM). There are also well known examples of the foliations
of compact spheres and tori, like Reeb and Kronecker foliations, which
also appear in some physical contexts. We see that indeed there are
many instances of the relevance of such (codimension-one) foliations
in physics, but the point is that these are, in some sense, trivial.
The invariants distinguishing certain equivalence classes of codimension-one
foliations of a compact manifold $M$ are the Godbillon-Vey (GV) invariants.
It was, soon after, shown by Thurston that GV invariant naturally
take values in $H^{3}(M,\mathbb{R})$. It distinguishes the cobordism
classes of codimension-one foliations of $M$. All the foliations
of compact manifolds mentioned above have vanishing GV invariant hence
are trivial from that point of view. Our main concern in this work
is the  physical meaning of  codimension-one foliations of the 3-sphere
(or other compact 3-manifolds) with non-vanishing GV class. This is
far less obvious that these have  any relevance to physics. To grasp
more concretely what are codimension-one foliations with ${\rm GV}\neq0$
let us turn to the connection with quantum physics. Even though not
all non-commutative $C^{\star}$-algebras are derived from some foliations,
almost all physically relevant cases are supposed to be. Namely the
factor I algebra is generated by the Reeb foliation whereas the factor
II by the Kronecker foliations of tori but both have vanishing GV
class. It is the factor III algebra which is generated by foliations
with ${\rm GV}\neq0$. So, on the one hand, foliations are classical
geometric structures whereas on the other hand, foliations are (extensions)
of quantum mechanical algebras of observables. This property makes
them particularly well-suited for exploring the regime of physical
theories which are formulated on manifolds and which aim towards the
description of quantum realm. This is the case of allmost all quantum
field theories.

So, we follow the recent proposition to use the codimension-one foliations
of certain 3-manifolds with ${\rm GV}\neq0$ for exploring the intermediate
region of classical and quantum field theories. On the other hand,
classical solutions of general relativity in diverse dimensions have
been analyzed from various perspectives for many years, and the instanton-like
configurations appeared as special in approaching a quantum theory
of gravity. They bear potential role in the determination of transition
amplitudes in quantum gravity being at the same time classical solutions
of GR and they are also well-suited towards exploring the intermediate
regime between classical and quantum gravity. In this paper we show
that, indeed, the relation between foliations with ${\rm GV}\neq0$
and gravitational instantons, can broaden our understanding of some
'quantum' limit of GR.

Gravitational instantons, in general, might be quite similar to the
case of instantons in gauge theory. Instantons are the self-dual finite
Euclidean-action configurations which minimize the action. The determination
of the complete set of instanton configurations is crucial for the
evaluation of the path integral of gauge theory in the semi-classical
approximation. The functional integral reduces then to an integral
over the instanton moduli space in each sector of the topological
charge $k$ %
\footnote{In pure Yang-Mills theory, the complete set of instantons, i.e self-dual
gauge fields of arbitrary topological charge $k$, is determined from
the matrix equations known as the Atiyah-Drinfeld-Hitchin-Manin (ADHM)
equations.%
}. Even though, the path integral over instanton moduli space is hardly
done directly, the localization technique, along with the tools from
non-commutative geometry, allow for the successful computations and
the results obtained in this way, perfectly match  Seiberg-Witten
theory (SW) \cite{Nekr2002e}.

So, extending the path integral over gravitation, one could expect
that 'gravitational instantons' play a similar substantial role in
quantum gravity. However, quantum gravity in 4 dimensions as the attempt
to quantize general relativity does not exist as any quantum field
theory built on the base of GR. This is mainly due to the appearance
of various divergences  which may suggest that Einstein's general
relativity in the present form is rather a low-energy or large-distance
approximation to some more fundamental theory. On the other hand 10-d
general relativity appears as classical approximation derived from
the $\beta$-function of superstring theory. Moreover, gravitons are
necessarily present in the spectra of the perturbative string theory.
These important features of higher dimensional theories if applicable
somehow in dimension 4 would be a crucial step in determining 4-d
QG. 

The Euclidean gravitational action is not positive-definite in general,
even for real positive-definite metrics, however, one can still evaluate
the functional integral by first looking for non-singular stationary
points of the action functional and expanding about them. These critical
points are finite action solutions to the vacuum Einstein equations
by self-dual or anti-self-dual complete, non-singular, and positive-definite
metrics \cite{EGH-1980}, so they are gravitational analogues of Yang-Mills
instantons, i.e. gravitational instantons \cite{Hawking-1977}. In
general it is shown \cite{GP-1979} that the self-dual or anti-self-dual
metrics are local minima of the action among metrics with zero scalar
curvature. Facing all these facts, we try to assign some gravitational
instantons to exotic $\mathbb{R}^{4}$ in such a way that this would
help evaluating  dominant contributions to the gravitational path
integral. As Witten argued in \cite{Witten85} the physical relevant
gravitational instantons are  represented precisely by exotic higher
dimensional spheres $S^{n},n=7,11$. This result was derived from
the description of the group of large diffeomorphisms of $S^{n-1}$
in terms of exotic smoothness on $S^{n}$. Since there are no exotic
spheres in dimensions $1,2,3,5,6,12$ and there exist, in particular,
exotic $S^{7}$ and $S^{11}$, Witten showed how these govern gravitational
anomalies in theories relevant to string theory in dimension 6 and
10. These exotic $S^{7}$ and $S^{11}$ serve as gravitational instantons.

One would like to recognize similarly the case of $S^{4}$ where one
expects  physical relevance as well. However, it is not known currently
whether an exotic $S^{4}$ exists (smooth 4-d Poincare conjecture).
Even though they exist, their connections with the large diffeomorphism
group of $S^{3}$, ${\rm Diff}(S^{3})$, would not follow the pattern
from higher dimensions (see e.g. \cite{Milnor1962}) employed by Witten.
Moreover, exotic $S^{4}$'s would be supported rather on Euclidean
asymptotically flat 4-geometries on $\mathbb{R}^{4}$. This means
that induced smoothness on $\mathbb{R}^{4}$ would be localized in
the exotic 4-disk. This, in turn, means that known small exotic $\mathbb{R}^{4}$s
are not of this type. Currently the smooth 4-d Poincare conjecture
is not resolved yet. On the other hand it is confirmed that there
exists a plenty of exotic smooth structures on Euclidean $\mathbb{R}^{4}$.
Hence, in this paper we analyze the case of exotic smooth $\mathbb{R}^{4}$s
grouped in a radial family (first described in \cite{DeMichFreedman1992})
and their relation with gravitational instantons. Even though, any
direct derivation of this kind of results is not possible at present,
the interesting pattern emerges via indirect reasoning based on quasi-modularity.

Recently it was shown that (cobordism classes of) codimension-one
foliations of $S^{3}$ with non-zero GV class distinguish between
different small exotic smooth $\mathbb{R}^{4}$s grouped in the so
called radial family %
\footnote{One can use also other closed compact 3-manifolds and their foliations.%
}. It was subsequently proposed to consider a special limit of (supersymmetric)
gauge theories formulated on exotic $\mathbb{R}^{4}$ from the radial
family, where geometric data of the foliations become dominant. In
the following we call it the \emph{foliated topological limit} (FTL)
which is not purely topological but rather contains expressions derived
from the gauge theory depend additionally on the characteristics of
the foliations (and this dependence is dominant). In this way the
gravitational corrections to Seiberg-Witten (SW) theory on the standard
$\mathbb{R}^{4}$ are derived from the foliated topological limit
of Seiberg-Witten theory formulated on exotic $\mathbb{R}^{4}$. The
driving principle is the quasi-modularity of the expressions: the
universal class of every codimension-one foliation acts via multiplying
by the quasi-modular Eisenstein 2-nd series, $E_{2}$, thus the characteristics
of the foliations are given by the polynomials of $E_{2}$ with modular
coefficients. The same kind of  expressions can be seen  among the
gravitational corrections to SW theory derived from superstring theory.
Some of them appear as instantons corrections within the instanton
calculus of Nekrasov. In general, the dependence on $E_{2}$ might
indicate  the role played by the codimension-one foliations.

Here we are looking for the corresponding foliated topological limit
of general relativity on the exotic $\mathbb{R}^{4}$. As usual, the
problem is our highly limited knowledge of exotic 4-metrics on $\mathbb{R}^{4}$
\footnote{In fact, we do not know any of them, but there exist plenty of smooth
metrics on every Riemannian smooth manifold.%
} and thus we are not able to find any solution with explicit exotic
metrics. However, quasi-modularity indicates  the possible role of
Bianchi IX geometry ${\cal M}_{3}$ as appearing in ${\cal M}_{3}\times\mathbb{R}$.
It is precisely the gravitational instanton where ${\cal M}_{3}$
is the fully anisotropic squashed 3-sphere. On the other hand $S^{3}\times\mathbb{R}$
is the topological end of $\mathbb{R}^{4}$ (and the smooth end of
the standard $\mathbb{R}^{4}$). In our case of GR on an exotic $\mathbb{R}^{4}$,
we find that a suitable 'geometric' end is usually based on fully
anisotropic squashed ${\rm sq}S^{3}$, i.e. ${\cal M}_{3}\times\mathbb{R}$.
One reason for this result is the metric on ${\rm sq}S^{3}\times\mathbb{R}$
written as function of the quasi-modular series $E_{2}$. In contrast, the
metrics on the fully symmetric $S^{3}\times\mathbb{R}$ and on partially
asymmetric spaces do not depend on $E_{2}$. Moreover, quasi-modularity
is expected in the special foliated topological limit of GR on exotic
$\mathbb{R}^{4}$s in the radial family. In this way foliations and
gravitational instantons meet at the semi-classical approximation
to QG. However, this gravitational instanton at the end of exotic
$\mathbb{R}^{4}$ is not presumably the part of the instanton solution
on this exotic 4-space. Rather, the instanton ${\cal M}_{3}\times\mathbb{R}$
gives the dominant contribution  to the path integral from exotic
$\mathbb{R}^{4}$ in the foliated topological limit of GR. Recall
that the technique of suitable 'ends' replacing the standard one,
was applied in the case of exotic $\mathbb{R}_{k}^{4},k=1,2,3...$
(from the fixed radial family). These were 'algebraic' ends, i.e.
$S_{k}^{3}\times\mathbb{R}$ which became relevant at the quantum
regime of the theory on the exotics. Then, the derivation of various
results from string theory was possible \cite{AssKrol2010ICM,AsselmKrol2011d,AsselmeyerKrol2011,AsselmeyerKrol2011b,AsselmKrol2011f}.
Even though, quasi-modularity characterizes foliations and exotic
4-spaces on more classical level, these expressions are derived from
characters of certain superconformal algebras \cite{AsselmKrol2012b}.

The connection with gravitational instantons, we believe, is fundamental
for 4-d QG and allows for the inclusion of effects of exotic $\mathbb{R}^{4}$s
into the path integral. Moreover, one finds again a surprising connection
of exotic 4-geometry with low energy interacting magnetic monopoles.
The point is that ${\cal M}_{3}\times\mathbb{R}$ is precisely the
geometry of the moduli space of $k=2$ low-energy BPS magnetic monopoles.
However, given the BPS condition one has charges determined by mass
or energy. This kind of gravitational sources can be found  due to
the change of the exotic smoothness on $\mathbb{R}^{4}$ to the standard
one. This idea is the explanation of the coincidence between magnetic
charge of Polyakov-'t Hooft monopoles and exotic 4-smoothness from
the fixed radial family observed for the limiting case of Dirac monopoles
in \cite{AsselmeyerKrol2009}. We close the paper with a discussion
of the potential physical significance of monopoles, hence Higgs field,
and the appearance of a  massive boson $A$, as generated from the
exotic 4-geometry on $\mathbb{R}^{4}$.

\section{Quasi-modularity from exotic $\mathbb{R}^4$s from the radial family}\label{q-m-2}
The first important step of our construction is the
relation of an exotic $\mathbb{R}^{4}$ to codimension-1 foliations
of $S^{3}$. To this end one fixes a radial family of small exotic
smooth $\mathbb{R}^{4}$. Let it be the Freedman-DeMichelis radial
family where our exotic $\mathbb{R}^{4}$s, $e$, belong to. \emph{Every
(small) exotic $\mathbb{R}^{4}$ is denoted by $e$ in the following.} However, beginning from the conjecture in Sec. \ref{GI-2}, $e$ denotes particular those exotic $\mathbb{R}^4$s from the radial family whose ends are \emph{modeled on AH instanton}.
Recall that every small exotic $\mathbb{R}^{4}$ belongs to some radial
family of such structures.

The radial family of exotic $\mathbb{R}^{4}$s, $R_{t}$, was discovered
and described firstly by Freedman and DeMichelis \cite{DeMichFreedman1992}
and is the main tool for  establishing  many relative results about
exotic $\mathbb{R}^{4}$s. This radial family consists of a continuum
of non-diffeomorphic smooth small $\mathbb{R}^{4}$s which are defined
on open subsets of standard $\mathbb{R}^{4}$ and are labeled by a
single radius $t\in{\rm CS}$ where CS is the standard Cantor set,
such that $t_{1},t_{2}\in{\rm CS}$ and $t_{1}<t_{2}$ then $\mathbb{R}_{t_{2}}^{4}\subset\mathbb{R}_{t_{1}}^{4}$.
It was first proved in \cite{AsselmeyerKrol2009} (but see also \cite{AsselmeyerKrol2011,AsselmKrol2011f})
that: \begin{theorem}\label{T1} Let us consider a radial family
$R_{t}$ of small exotic $\mathbb{R}_{t}^{4}$ with radii $\rho$,
where $t=1-\frac{1}{\rho}$ $\rho\in{\rm CS}\subset[0,1]$,  induced
from the non-product $h$-cobordism $W$ between $M$ and $M_{0}$.
Then, the radial family $R_{t}$ determines a family of codimension-one
foliations of $S^{3}$ with Godbillon-Vey invariant $\rho^{2}$. Furthermore,
given two exotic spaces $R_{t}$ and $R_{s}$, homeomorphic but non-diffeomorphic
to each other (and so $t\neq s$), the two corresponding codimension-one
foliations of $S^{3}$ are non-cobordant to each other. \end{theorem}
The following important, though direct, consequence holds true: \begin{corollary}
Every class in $H^{3}(S^{3},\mathbb{R})$ induces a small exotic $\mathbb{R}^{4}$
where $S^{3}$ lies at the boundary of some compact subset of $\mathbb{R}^{4}$.
\end{corollary} Next, given a codimension-1 foliation of a manifold
$N$, with ${\rm GV}\neq0$, one assigns a universal GV class to it.
This construction is due to Connes-Moscovici: then the universal GV
class is  represented by the cocycle in the cyclic cohomologies of
the special universal Hopf algebra ${\cal H}_{1}$. A natural action
of ${\cal H}_{1}$ on the crossed product ${\cal M}\ltimes{\rm GL}^{+}(2,\mathbb{Q})$,
of the ring of modular forms ${\cal M}$ and the matrix group ${\rm GL}^{+}(2,\mathbb{Q})$,
is determined. Thus, given the Theorem \ref{T1} the following crucial
result  emerges \cite{AsselmKrol2012b}: \begin{theorem}[Th. 6.6, \cite{AsselmKrol2012b}]
For the DeMichelis-Freedman radial family of exotic $\mathbb{R}^{4}$'s
the exotic smoothness structures of the members determine a non-holomorphic
canonical deformation of the modular functions from ${\cal M}$. This
deformation is determined by the action of the Hopf algebra ${\cal H}_{1}$
on the crossed product ${\cal M}\rtimes{\rm GL}^{+}(2,\mathbb{Q})$
where the GV class of the codimension-one foliations of $\Sigma$
is interpreted as the cyclic cocycle of the ${\cal H}_{1}$. \end{theorem}
From the Connes-Moscovici construction follows that the universal
GV class acts via multiplication by $E_{2}$. In that way monomials
of Eisenstein series $f\cdot E_{2}$, and subsequently polynomials,
with modular coefficients appear. The appearance of these expressions
indicates that the special, half topological half geometric, limit
of a theory is presumably approached and, in dimension 4, could be
related, under certain conditions, with exotic 4-smoothness on $\mathbb{R}^{4}$.

\section{Foliated topological limit of GR on $e$}
Let us consider a smooth connected 4-manifold $M$, compact or not,
and some closed 3-d smooth manifold $N$ such that $N\subset M$ is
a smooth submanifold. Let us suppose that there exists another smooth
structure on $M$, $\tilde{M}$, and that now, $N\subset\tilde{M}$
is only a topological submanifold. Let us consider the codimension-1
foliations ${\cal F}_{N}$ of $N$, with non-zero Godbillon-Vey invariants,
${\rm GV}_{N}\in H^{3}(N,\mathbb{R})$. The solution of a sourceless
Einstein equations (EE) on $M$ determines a metric $g_{M}$ on $M$
(a Ricci-flat metric). Suppose that the smooth structure $\tilde{M}$
is related with the GV class of the foliations ${\cal F}_{N}$s in
an invariant manner which means that $\tilde{M}$ can be determined
from ${\rm GV}({\cal F}_{N})\neq0$. We define the foliated topological
limit of GR in this case, as: \begin{definition}\label{def-1} The
foliated topological limit (FTL) of GR on exotic $\tilde{M}$ is given
by: 
\begin{itemize}
\item i. a solution of EE on $\tilde{M}$ - a smooth metric $\tilde{g}$
on $\tilde{M}$ such that 
\item ii. $\tilde{g}$ is the function of Eisenstein 2-nd quasi-modular
series and $\tilde{g}$ is derivable from the quasi-modular properties
of $E_{2}$. 
\end{itemize}
\end{definition} Let us recall a similar construction in the case
of supersymmetric Yang-Mills theory on exotic $\mathbb{R}^{4}$ \cite{AsselmKrol2012c}.
For the low energy effective theory to the ${\cal N}=2$ YM, i.e.
Seiberg-Witten theory on exotic $\mathbb{R}^{4}$, the FTL of the
theory was defined similarly, namely, as the regime where the existing
corrections to the Langrangian on flat standard $\mathbb{R}^{4}$
 are polynomials in the Eisenstein 2-nd series with modular coefficients.
The main point in deriving these results was the existence of a single
prepotential for SW theory so that the corrections were basically
the corrections to the prepotential. However, it was obtained for
${\cal N}=2$ supersymmetry which guarantees  the exsitence of the
prepotential at all. In our case of GR we do not make any use of supersymmetry
and try to obtain a similar dependence on $E_{2}$ though directly
at the level of metric. Since, we have the limited possibility of
dealing with explicit exotic metrics on $\mathbb{R}^{4}$, the FTL
defined above serves as a general heuristic rule for grasping the
dependence of GR on the foliations which determine the fake $\mathbb{R}^{4}$
(see, Sec. \ref{q-m-2}).

Before going on, let us comment on the general meaning of the FTL
limit which also motivates our next steps. The FTL can be understood
as a kind of geometry/geometry duality in field theory and gravity.
Namely, given a background manifold $M$ the change of smoothness
induces some relevant changes of the geometry. But now the new geometry
relies on the foliations of $N$ because the exotic smoothness of
$M$ depends on it. Moreover, the new geometry of exotic $\tilde{M}$
places itself in the non-perturbative regime of the theories, hence,
the connection with monopoles and instantons is rather natural. In
FTL one searches for a formulation of the theories on $\tilde{M}$
which would recover the characteristics of the foliations as the parts
of the Lagrangians or the correlation functions. In the case of classical
field theory, like GR, one recovers solutions of the theory as characteristics
of the foliations. This is analogous to taking a 'topological' twist
in some (supersymmetric) QFTs, or in theories of gravity. However,
in our case of FTL, the characteristics  respects rather the inherently
non-commutative geometry of the foliations. In this way quantumness
is already written in the classical solutions of the theory and is
connected with a dual 4-geometry. Note that the geometry of instantons
has also classical and as well  quantum meaning in semiclassical and
Euclidean QG. Searching for gravitational instanton-like geometries
as assigned to dual exotic 4-geometry seems legitimate from that point
of view.

\section{Gravitational instanton from quasi-modularity}\label{Gr-I}
Our next task is to find 4-d metrics which would be
generated from quasi-modular expressions, in order to fulfill the
requirements of Def. \ref{def-1}. Thus, let us begin with $y(z)=i\pi E_{2}(z),z\in\mathbb{C}$.
One additionally derives: $y'(z)=\frac{i\pi}{6}(E_{2}^{2}-E_{4})$
and $y''(z)=-\frac{i\pi^{3}}{18}(E_{2}^{3}-3E_{2}E_{4}+2E_{6})$ where
$E_{2},E_{4},E_{6}$ are Eisenstein series \cite{Zagier2008}. With
$\vartheta_{2},\vartheta_{3},\vartheta_{4}$ the Jacobi theta-functions
\cite{Zagier2008}, let us introduce new variables $\omega^{1},\omega^{2},\omega^{3}$
as: \begin{equation}
\begin{array}{c}
\omega^{1}=\frac{\pi}{6i}(E_{2}-\vartheta_{2}^{4}-\vartheta_{3}^{4})\\[5pt]
\omega^{2}=\frac{\pi}{6i}(E_{2}-\vartheta_{3}^{4}-\vartheta_{4}^{4})\\[5pt]
\omega^{3}=\frac{\pi}{6i}(E_{2}+\vartheta_{2}^{4}-\vartheta_{4}^{4})\;,\end{array}\label{o-1}\end{equation}
so that $y=-2(\omega^{1}+\omega^{2}+\omega^{3})$ and $y'=2(\omega^{1}\omega^{2}+\omega^{2}\omega^{3}+\omega^{3}\omega^{1})$,
$y''=-12\omega^{1}\omega^{2}\omega^{3}$ and the Jacobian of the coordinate
change reads: $J=(\omega^{1}-\omega^{2})(\omega^{2}-\omega^{3})(\omega^{3}-\omega^{1})$,
which is non-zero for $\omega^{1}\neq\omega^{2}\neq\omega^{3}$.

Now, let us consider the so called Darboux-Halphen (DH) system of
differential equations on $\gamma^{i}(z),z\in\mathbb{C},i=1,2,3$:
\begin{equation}
\begin{array}{c}
\frac{{\rm d}\gamma^{1}}{{\rm d}z}=\gamma^{2}\gamma^{3}-\gamma^{1}(\gamma^{2}+\gamma^{3})\\[5pt]
\frac{{\rm d}\gamma^{2}}{{\rm d}z}=\gamma^{2}\gamma^{3}-\gamma^{1}(\gamma^{2}+\gamma^{3})\\[5pt]
\frac{{\rm d}\gamma^{3}}{{\rm d}z}=\gamma^{2}\gamma^{3}-\gamma^{1}(\gamma^{2}+\gamma^{3})\;.\end{array}\label{o-2}\end{equation}
It is known that unlike the other cases, the fully anisotropic one,
i.e. $\gamma^{1}\neq\gamma^{2}\neq\gamma^{3}$, does not allow for
algebraic integrals. The solutions are, however, expressible in terms
of quasi-modular series $E_{2}$. Namely, given the base for quasi-modular
forms of weight 2, i.e. ${\cal E}^{i},i=1,2,3$, the fully anisotropic
solutions of DH system (\ref{o-2}), read: \begin{equation}
\gamma^{i}(z)=-\frac{1}{2}\frac{{\rm d}}{{\rm d}z}{\rm log}{\cal E}^{i}(z),i=1,2,3\,.\label{l-1}\end{equation}
Defining $\lambda=\frac{\gamma^{1}-\gamma^{3}}{\gamma^{1}-\gamma^{2}}$
one expresses ${\cal E}^{i},i=1,2,3$ as: \begin{equation}
{\cal E}^{1}=\frac{1}{\lambda}\frac{{\rm d}\lambda}{{\rm d}z},\;{\cal E}^{2}=\frac{1}{\lambda-1}\frac{{\rm d}\lambda}{{\rm d}z},\;{\cal E}^{3}=\frac{1}{\lambda(\lambda-1)}\frac{{\rm d}\lambda}{{\rm d}z}\;.\end{equation}
Now, the choice of $\lambda=\frac{\vartheta_{2}^{4}}{\vartheta_{3}^{4}}$
and calculating ${\cal E}^{i}$ result in: \[
{\cal E}^{1}=i\pi\vartheta_{4}^{4},\;{\cal E}^{2}=-i\pi\vartheta_{2}^{4},\;{\cal E}^{3}=-i\pi\vartheta_{3}^{4}\,.\]
Equivalently, one derives $\gamma^{i}$ from (\ref{l-1}) as: \begin{equation}
\gamma^{1}=\frac{\pi}{6i}(E_{2}-\vartheta_{2}^{4}-\vartheta_{3}^{4}),\;\gamma^{2}=\frac{\pi}{6i}(E_{2}-\vartheta_{3}^{4}-\vartheta_{4}^{4}),\;\gamma^{3}=\frac{\pi}{6i}(E_{2}+\vartheta_{2}^{4}-\vartheta_{4}^{4})\,.\label{l-2}\end{equation}
However, these are precisely the expressions for $\omega^{i}$ as
in (\ref{o-1}), i.e. $\omega^{i}=\gamma^{i},i=1,2,3$. Starting from
$i\pi E_{2}$ we arrive at the solutions of (\ref{o-2}). Now we will
see that solving Eqs. \ref{o-2} is nothing but finding certain 4-d
metric which is a gravitational instanton. First, we are going to
show that the requirement for self-duality in the GR equations on
certain 4-manifold is recapitulated precisely in the system (\ref{o-2})
of DH equations. Let ${\cal M}_{3}$ be 3-space of Bianchi IX type,
i.e. the Killing vectors $\xi_{i},i=1,2,3$ form the $SU(2)$ algebra:
\[
[\xi_{i},\xi_{j}]=c_{\; jk}^{i}\xi_{k}\]
and the Maurer-Cartan 1-forms, $\sigma^{i},i=1,2,3$ for this Bianchi
group fulfill: \[
{\rm d}\sigma^{i}=\frac{1}{2}c_{\; jk}^{i}\sigma^{j}\wedge\sigma^{k}\,.\]
In terms of the Euler angles, $0\leq\alpha\leq\pi,0\leq\beta\leq2\pi,0\leq\psi\leq4\pi$,
these 1-forms read: \begin{equation}
\begin{array}{c}
\sigma^{1}=\sin{\alpha}\sin{\psi}{\rm d}\beta+\cos{\psi}{\rm d}\alpha\\[5pt]
\sigma^{2}=\sin{\alpha}\cos{\psi}{\rm d}\beta-\sin{\psi}{\rm d}\alpha\\[5pt]
\sigma^{3}=\cos{\alpha}{\rm d}\beta+{\rm d}\psi\;.\end{array}\label{o-4}\end{equation}
A metric ${\rm d}s_{3}^{2}=g_{ij}\sigma^{i}\sigma^{j}$ on ${\cal M}_{3}$
can be always diagonalized. Let us take  our 4-d manifold ${\cal M}_{3}\times\mathbb{R}$.
General metric on it reads: \[
{\rm d}s^{2}={\rm d}t'^{2}+g_{ij}(t')\sigma^{i}\sigma^{j}\,.\]
However, this metric can be always diagonalized too, and with the
use of arbitrary functions $\Theta^{i}(t'),i=1,2,3$, such that ${\rm d}t'=\sqrt{\Theta^{1}\Theta^{2}\Theta^{3}}{\rm d}t$,
a general metric on ${\cal M}_{3}\times\mathbb{R}$ becomes: \begin{equation}
{\rm d}s^{2}=\Theta^{1}\Theta^{2}\Theta^{3}{\rm d}t^{2}+\frac{\Theta^{2}\Theta^{3}}{\Theta^{1}}(\sigma^{1})^{2}+\frac{\Theta^{1}\Theta^{3}}{\Theta^{2}}(\sigma^{2})^{2}+\frac{\Theta^{1}\Theta^{2}}{\Theta^{3}}(\sigma^{3})^{2}\,.\label{o-5}\end{equation}

Now, as usual in the 4-d Cartan formalism, the basis of 1-forms on
a 4-manifold $M$ is given by $\{\theta^{a}\}$. Then the corresponding
connection 1-forms $\omega_{\; b}^{a}=\Gamma_{\; bc}^{a}\theta^{c},a,b,c=1,2,3,4$,
the curvature 2-form reads: \begin{equation}
{\cal R}_{\; b}^{a}={\rm d}\omega_{\; b}^{a}+\omega_{\; c}^{a}\wedge\omega_{\; b}^{c}=\frac{1}{2}R_{\; bcd}^{a}\theta^{c}\wedge\theta^{d}\label{o}\end{equation}
where $R_{\; bcd}^{a}$ are the components of the Riemann tensor.
Let us now decompose the Riemann curvature ${\cal R}_{ab}$ and the
connection $\omega_{ab}$ into self-dual and anti-self-dual parts
with respect to the duality defined by: \begin{equation}
\begin{array}{c}
\tilde{\omega}_{\; b}^{a}=\frac{1}{2}\varepsilon_{\; bc}^{a\;\; d}\omega_{\; d}^{c}\\[5pt]
\tilde{{\cal R}}_{\; b}^{a}=\frac{1}{2}\varepsilon_{\; bc}^{a\;\; d}{\cal R}_{\; d}^{c},\end{array}\label{o-6}\end{equation}
and with respect to the factorization $SO(4)$ into $SO(3)\times SO(3)$.
Then, writing for $i,j,k=1,2,3$: \begin{equation}
\begin{array}{c}
s_{i}=\frac{1}{2}(\omega_{0i}+\frac{1}{2}\varepsilon_{ijk}\omega^{jk}),\;\;\;\;\; a_{i}=\frac{1}{2}(\omega_{0i}-\frac{1}{2}\varepsilon_{ijk}\omega^{jk})\;,\\[5pt]
S_{i}=\frac{1}{2}({\cal R}_{0i}+\frac{1}{2}\varepsilon_{ijk}{\cal R}^{jk})\;\;\;\;\; A_{i}=\frac{1}{2}({\cal R}_{0i}-\frac{1}{2}\varepsilon_{ijk}{\cal R}^{jk}),\end{array}\label{o-7}\end{equation}
one has $s_{i}=\tilde{s}_{i}$, $S_{i}=\tilde{S}_{i}$ and $a_{i}=-\tilde{a}_{i}$,
$A_{i}=-\tilde{A}_{i}$ under the duality (\ref{o-6}). Moreover,
${\cal R}_{\; b}^{a}$ in (\ref{o}) is given by: \begin{equation}
\begin{array}{c}
S_{i}={\rm d}s_{i}-\varepsilon_{ijk}s^{j}\wedge s^{k}\\[5pt]
A_{i}={\rm d}a_{i}+\varepsilon_{ijk}a^{j}\wedge a^{k}\;,\end{array}\label{o-8}\end{equation}
so that $s_{i},S_{i}$ are self-dual components of $\omega$ and ${\cal R}_{\; b}^{a}$,
while $a_{i},A_{i}$ are anti-self-dual their components. In this
way the self-duality equations of GR read: \begin{equation}
A_{i}={\rm d}a_{i}+\varepsilon_{ijk}a^{j}\wedge a^{k}=0\,.\label{o-9}\end{equation}
In terms of the diagonalizing functions $\Theta^{i},i=1,2,3$ for
our case of $M={\cal M}_{3}\times\mathbb{R}$, we have: \begin{equation}
a_{i}=\frac{1}{4\sqrt{\Theta^{1}\Theta^{2}\Theta^{3}}}\left[\frac{1}{\Theta^{i}}\left(\frac{{\rm d}\Theta^{i}}{{\rm d}t}-\Theta^{j}\Theta^{k}\right)-\frac{1}{\Theta^{j}}\left(\frac{{\rm d}\Theta^{j}}{{\rm d}t}-\Theta^{k}\Theta^{i}\right)-\frac{1}{\Theta^{k}}\left(\frac{{\rm d}\Theta^{k}}{{\rm d}t}-\Theta^{i}\Theta^{j}\right)\right]\,.\label{o-10}\end{equation}
One solves (\ref{o-9}) for non-zero $a_{i}$ yielding $a_{i}=\frac{1}{2}\delta_{ij}\sigma^{j}$,
which together with (\ref{o-10}) gives the following system of equations:
\begin{equation}
\begin{array}{c}
\frac{{\rm d}\Theta^{1}}{{\rm d}t}=\Theta^{2}\Theta^{3}-\Theta^{1}(\Theta^{2}+\Theta^{3})\\[5pt]
\frac{{\rm d}\Theta^{2}}{{\rm d}t}=\Theta^{3}\Theta^{1}-\Theta^{2}(\Theta^{3}+\Theta^{1})\\[5pt]
\frac{{\rm d}\Theta^{3}}{{\rm d}t}=\Theta^{1}\Theta^{2}-\Theta^{3}(\Theta^{1}+\Theta^{2})\end{array}\label{o-11}\end{equation}
which is nothing but the real line version of the Darboux-Halphen
equations (\ref{o-2}). To retrieve real solutions $\Theta^{i}(t)$
from the complex ones $\gamma^{i}(z)$ as in (\ref{l-1}), one simply
takes: \begin{equation}
\Theta^{k}(t)=i\gamma^{k}(it)=-\frac{1}{2}\frac{{\rm d}}{{\rm d}t}{\rm log}{\cal E}^{k}(it),k=1,2,3\,.\label{o-12}\end{equation}
In this way we have a metric on ${\cal M}_{3}\times\mathbb{R}$ as
in (\ref{o-5}) whose coefficients depend on $E_{2}$ as in (\ref{o-1})
and which were, in fact, determined from the quasi-modularity of $E_{2}$.
To complete the story we need to show that this metric is the gravitational
instanton. But, its self-duality was imposed by (\ref{o-9}) and (\ref{o-10})
and this condition is in fact realized in the DH eqs. (\ref{o-11}).
This solution has a long story and was also obtained by Atiyah and
Hitchin in the context of monopoles \cite{AH-1985,AH-1988}, so it
is called Atiyah-Hitchin (AH) instanton.

\section{Gravitational instanton and magnetic monopoles from exotic $\mathbb{R}^4$}\label{GI-2}
In this section we try to go further and relate the gravitational
instanton on ${\cal M}_{3}\times\mathbb{R}$, obtained in the previous
section, with the 'effective' geometry of the end of the exotic $\mathbb{R}^{4}$.

$SU(2)$ YM theory can be formulated on every smooth, compact or not,
4-manifold. Moreover, it is known that the YM $SU(2)$ instanton exists
on every compact oriented 4-d smooth manifold \cite{Taubes-1984}.
It was believed that it exists also on open oriented 4-d manifolds.
However, the counterexample was constructed in \cite{Tsukamoto2010}
showing that the infinite connected sum of the complex projective
space, $(CP^{2})^{\#\mathbb{Z}}$, does not allow for any (non-flat)
YM instanton solution. There is also a close relation between YM and
gravitational instantons in the sense that a large class of YM instantons
is solved at the same time by a gravitational instanton \cite{Oh-2011}.

Thus, currently, we can not decide definitely whether gravitational
instantons exist on exotic $\mathbb{R}^{4}$s. Instead, we propose
to associate one with these curved open 4-manifolds such that, in
some limit, these instantons have dominant contributions to the path
integral. So we perform two deformations, one when the standard smooth
structure on $\mathbb{R}^{4}$ is changed to $e$, and the second
within GR on $e$ by taking FTL. As usual we fix the radial family
which the exotic $e$ belongs to. As so, the (cobordant classes of)
codimension-1 foliations of $S^{3}$ with non-zero GV, label exotics
$e$'s from the family. In fact, given ${\rm GV}\neq0$ one determines
the (cobordant classes of) codimension-1 foliations of $S^{3}$ from
the radial family. Taking the FTL limit of a theory on $e$ is similar
to the topological invariants of the manifolds obtained as the correlation
functions in the topological twist of a quantum field theory. Here
we have a rather quasi-topological twist of a theory where the expressions
depend on $E_{2}$ hence on the class of the foliation. So, the quasi-modular
character of metrics in the FTL limit of GR on $e$ is expected. This
is also an indication that the instanton, given in (\ref{o-5}) and
(\ref{o-1}), might be relevant here.

Let us take a closer look at the geometry of this instanton. The case
of the DH equations makes use of $\Theta^{1}\neq\Theta^{2}\neq\Theta^{3}$.
The partially anisotropic case, as e.g., $\Theta^{1}=\Theta^{2}$,
and the fully symmetric one, $\Theta^{1}=\Theta^{2}=\Theta^{3}$,
do not require any reference to quasi-modular expressions. These cases
are solved by algebraic means and are related with other gravitational
instantons, like Eguchi-Hanson or Taub-NUT ones. However, the geometry
of the fully symmetric case allows for isometries given by left-invariant
killing vectors (\ref{o-4}) creating one copy of $SU(2)$, and by
the other 3 right invariant killing vectors spanning the other $SU(2)$.
The isometries are, thus, $SU(2)\times SU(2)$, the spin group $Spin(4)$
of the Euclidean space. 4-d geometry resulting from these symmetries
is $S^{3}\times\mathbb{R}$. This is the topological, and smooth,
end of the standard $\mathbb{R}^{4}$. The partially anisotropic case
reduces the isometries to $SU(2)$ left invariant, as in (\ref{o-4}),
and breaks the right invariant $SU(2)$ to $U(1)$, so the result
is $SU(2)\times U(1)$. The geometry of this case is based on the
partially anisotropic (partially squashed) sphere i.e., ${\rm psq}S^{3}\times\mathbb{R}$.

Finally, the fully anisotropic, and quasi-modular, case is given by
the $SU(2)$ isometries, and its geometry is based on the fully anisotropic
3-sphere, i.e. ${\rm sq}S^{3}\times\mathbb{R}$. This 'subtle' squashing
of $S^{3}$ is the reason that one can not solve the self-duality
equations algebraically, and  quasi-modularity has to intervene. We
have to remember that the 3-d metrics of the 3-spheres depend on the
$t$ component and this $t$-dependence of the metrics is quasi-modular.
Thus, the deformation of $S^{3}$ (anisotropic squashing) and the
requirement of self-duality on ${\rm sq}S^{3}\times\mathbb{R}$ automatically
gives rise to the quasi-modular dependence on $t$. Without squashing
the self-duality alone does not produce any quasi-modular expressions.

We refer to the squashing of a 3-sphere as associated to the change
of the smoothness structure on $\mathbb{R}^{4}$. However, the effects
due to the change of smoothness are on the topological counterpart
of $S^{3}$, since it and the squashed ${\rm sq}S^{3}$ are not smoothly
embedded into exotic $\mathbb{R}^{4}$. Moreover, the resulting embedding
of $S^{3}$ should be based  on some wild embeddings. However, the
effects are understood in the sense of dominant contributions to the
path integral. Any direct calculation on $e$ is not possible at present.
We are rather approaching  qualitatively the change of the standard
smoothness of $\mathbb{R}^{4}$ into an exotic $e$ from the fixed
radial family. The smoothness of $e$ can not be localized into a
smooth 4-disk, so the exotic geometry deforms the standard end $S^{3}\times\mathbb{R}$.
The exotic end of $e$ is not a smooth $S^{3}\times\mathbb{R}$ nor
it is a smooth ${\rm sq}S^{3}\times\mathbb{R}$. We do not know how
precisely the geometry of this deformation looks like, however in
the FTL of GR on $e$ the quasi-modularity should be dominant. Moreover,
from the point of view of the path integral this should be assigned
to the solution minimizing the action. Both conditions are met on
the 'squashed end' ${\rm sq}S^{3}\times\mathbb{R}$ of $e$ which
is the case of the fully anisotropic AH instanton. More arguments
of purely mathematical character will be presented separately, here
we rather focus  on possible physics behind it. So, let us formulate
this relation as the following conjecture: 
\begin{conjecture} There
exists an exotic $\mathbb{R}^{4}$, $e$, belonging to the DeMichelis-Freedman
radial family such that:

The foliated topological limit of general relativity on $e$ is represented
by the geometric end given by the Atiyah-Hitchin instanton in the
sense that the contribution from this gravitational instanton approaches
the dominant contribution from $e$ into the gravitational Euclidean
path integral. \end{conjecture}\label{conj} Note that when smoothness
of $\mathbb{R}^{4}$ is fixed to the standard one, the foliations
have vanishing GV class, hence the whole quasimodularity disappears
(from the Connes-Moscovici construction). This corresponds to ${\rm sq}S^{3}\times\mathbb{R}\to S^{3}\times\mathbb{R}$.
However, beginning with the standard $\mathbb{R}^{4}$ and its standard
end $S^{3}\times\mathbb{R}$ and searching for its deformation such
that the dependence on $E_{2}$ becomes apparent, one arrives uniquely
at the stable geometry of the fully anisotropic $S^{3}$, i.e. ${\cal M}_{3}\times\mathbb{R}$.

Now let us comment on the contributions to the path integral of Euclidean
quantum gravity (EQG). Usually, the calculation of this path integral
is the integration over different classes of positive definite metrics
$g_{\mu\nu}$ on $e$, i.e.: \[
Z_{e}=\int D[g_{\mu\nu}]e^{-S[g_{\mu\nu},e]}\]
where the action can read: \[
S[g_{\mu\nu}]=-\frac{1}{16\pi}\int_{e}(R-2\Lambda)\sqrt{g}{\rm d}^{4}x-\frac{1}{8\pi}\int_{\partial e}K\sqrt{h}{\rm d}^{3}x=-\frac{1}{16\pi}\int_{e}(R-2\Lambda)\sqrt{g}{\rm d}^{4}x\,.\]
Given the decomposition of the standard $\mathbb{R}^{4}$ as $\mathbb{R}^{4}=D^{4}\cup S^{3}\times\mathbb{R}$
the path integral on the standard $\mathbb{R}^{4}$ reads \[
Z_{\mathbb{R}^{4}}=\int D[h]\left(\int D[g']e^{-S[g',h,D^{4}]}\int D[g'']e^{-S[g'',h,\mathbb{R}^{4}\setminus D^{4}]}\right)\]
which is not exotic-smooth valid, since $(S^{3},h)=(\partial D^{4},h)$
is not smoothly embedded in $e$ any longer, and $h$ is a 3-metric
on $S^{3}$. Now, we change the smooth structure to $e$. The decomposition
above do not respects the exotic smoothness structure. However, in
the exotic $e$ case, we still can speak about the (exotic) end, ${\rm end}(e)$,
such that $e$ is decomposed as $N^{4}\cup{\rm end}(e)$. In the FTL
of GR on $e$ the conjecture states that $Z_{e}^{FTL}$ is approximated
by the $\int D^{FTL}[g]e^{-S[g,{\cal M}_{3}\times\mathbb{R}]}$ in
the sense that the leading contribution from the ${\rm end}(e)$ in
FTL of GR on $e$, is calculated \emph{locally} from the AH instanton
in the sense given below.

Let $g_{\mu\nu}^{0}$ be a flat metric on $\mathbb{R}^{4}$ with its
restriction $g_{\mu\nu|S^{3}\times\mathbb{R}}^{0}$ to the standard
end $S^{3}\times\mathbb{R}$. Next, we are changing the smooth structure
on $\mathbb{R}^{4}$ to $e$. A non-flat metric on $e$ is $g_{\mu\nu}^{e}$
which is non smooth in the standard $\mathbb{R}^{4}$. Working in
the standard smoothness there is the difference $\tilde{\Phi}_{\mu\nu}$
(non-standard smooth) between the metrics, such that $g_{\mu\nu}^{e}=g_{\mu\nu}^{0}+\tilde{\Phi}_{\mu\nu}$
and $\tilde{\Phi}_{\mu\nu}$ is still continuous. Let us approximate
it by the sequence of standard smooth functions $\Phi_{\mu\nu}^{\epsilon}$
converging to $\tilde{\Phi}_{\mu\nu}$. Now $g_{\mu\nu}^{\epsilon}=g_{\mu\nu}^{0}+\Phi_{\mu\nu}^{\epsilon}$
is standard smooth. However, these $\Phi_{\mu\nu}^{\epsilon}$, 'sufficiently
well' (truely) approximated  exotic non-smooth corrections can not
be small perturbations everywhere. They are rather global effects
of the exotic smooth $e$.

In the semi-classical QG approximation we consider small $\phi_{\mu\nu}^{\epsilon}$
corrections as (quantum) perturbations which can be taken as $c$-number
fields.  Then, one can expand the path integral over the base geometry.
This base geometry is the saddle-point of the classical action and
is given by the complete, finite-action, metric, i.e. gravitational
instanton. For the case of the AH gravitational instanton, we have
a local expansion around it: \[
Z^{{\rm std}}[g_{\mu\nu}^{AH},{\cal M}_{3}\times\mathbb{R}]\approx e^{-S(g_{\mu\nu}^{AH})}\int D\phi^{\dag}D\phi\exp{[-\frac{1}{2}\int\sqrt{g}{\rm d}\tau{\rm d}^{3}x\phi^{\dag}{\cal A}(g^{AH})\phi]}=\exp{\{-S(g^{AH})-\frac{1}{2}\ln{\det}[{\cal A}(g^{AH})]\}}\]
where ${\cal A}$ is the quantum operator representing the corrections.
The claim in the conjecture states that the above expression is the
correct approximation for the path integral for GR on $e$ in FTL.
In fact, this is the approximation for the evaluation of the path
integral on the exotic end of $e$ in FTL, i.e. \begin{equation}
Z^{e}[g^{FTL},{\rm end}(e)]\approx Z^{{\rm std}}[g_{\mu\nu}^{AH},{\cal M}_{3}\times\mathbb{R}]\,.\label{PI-1}\end{equation}
Certainly, we do not know an explicit shape of the operator ${\cal A}$
but  we state its existence and that ${\cal A}$ depends on the particularities
of FTL of GR on $e$. Note that one can consider also other fields
on $e$ when path integral is approached. This is due to the fact
that the change of the smoothness modifies every smooth object on
$\mathbb{R}^{4}$.

One can say that the functional measure in FTL respects the universal
class of the codimension-1 foliations of $S^{3}$ in a sense that
the resulting expression depends also on the class of the foliation.
Suppose that FTL is attained for GR on $e$. Then, the relevant metrics
in this limit depend on $E_{2}$ and these metrics contribute to the
path integral. The conjecture states that $Z_{e}^{FTL}$ is approximated
by the contribution from the AH instanton, even though it is not any
exotic geometry at the end of the exotic $e$. As we will see the
AH geometry rather emerges  as a kind of 'symmetry breaking' in exotic
geometry by a scalar field $\phi$. This will enforce the $U(1)$
direction in the $SU(2)$ isospace and the quasi-modular direction
($t$) in 4-space.

To grasp contributions of other exotic $\mathbb{R}^{4}$s from this
radial family into the path integral, one should additionally respect
the dependence on the particularities of the solutions of the DH system
and/or on the non-universal GV classes of the foliations distinguishing
between the members of the radial family. The attempt to calculate
directly the gravitational path integral on exotic $\mathbb{R}^{4}$s
from the radial family, was performed recently \cite{Asselm-Krol-2011}.
It would be interesting to compare both approaches and check whether
the results agree in the case of a fixed radial family, where dependence
on the GV class is expected. We will address this possibility separately.

Thus, when we deal with gravity path integral in the FTL of GR on
$e$, the dominant contributions are derivable from the AH instanton
metric, and we, then, say that \emph{$e$ has geometric end modeled
on the AH instanton}.

The profound physical consequences can be drawn from the existence
of this exotic $\mathbb{R}^{4}$ with the geometric end modeled on
the Atiyah-Hitchin instanton in the FTL limit of GR. The relation
with magnetic monopoles is one consequence. Let us recall the following
result which was obtained originally in \cite{AsselmeyerKrol2009}:

\emph{Some small, exotic smooth structures on $\mathbb{R}^{4}$ can
act as sources of magnetic field, i.e. monopoles, in spacetime. Electric
charge in spacetime has to be quantized, provided some region has
this small exotic smoothness.}

This result was obtained due to the algebraic coincidence of universal
classes (cobordism) of the foliations and of Dirac monopoles. Here
we can go much further. We will show that the exotic $e$ as in the
conjecture, is related analytically with magnetic monopoles of Polyakov-'t Hooft
type, and this $e$ serves as the example of the exotic 4-region.
Moreover, there is the mechanism for the geometric generation of the
scalar Higgs field in 4-spacetime.

How is it possible that 4-geometry generates any monopole configuration
at all? Such possibility is encoded in the BPS condition. Let us consider
the non-flat geometry on exotic $e$. According to early results on
exotic $\mathbb{R}^{4}$s \cite{Bra:94b,Ass:96,Sladkowski2001,AsselBrans2011}
suppose that $e$ generates a gravitational source $T_{\mu\nu}\neq0$
which becomes apparent when the exotic smoothness of $e$ is changed
into the standard one. Thus, $e$ generates the non-zero source $T_{\mu\nu}$,
though Einstein equations on $e$ would be still sourceless in the
exotic structure. The non-trivial source $T_{\mu\nu}$ is, as usual,
connected with some distribution of matter and energy in spacetime.
Let us assume additionally that the matter in the static limit is
related with particles (classical or quantum) with non-zero masses
$m$. According to the above discussion it is natural to consider
$m$ as the mass of magnetic monopoles. However, the BPS condition
relates the charges (of dyons or monopoles) with its masses. So, in
the standard smooth structure of $\mathbb{R}^{4}$ the particles with
mass $m$ bear the monopole (dyon) charge. Moreover, the BPS condition
is crucial for determining the geometry ${\rm sq}S^{3}\times\mathbb{R}$
as the geometry of the moduli space with $k=2$. This geometry, in
turn, is derivable from the exotic $e$ due to quasi-modularity and
appears as the crucial player in evaluation of the gravitational path
integral on $e$. Note that, one can solve analytically the Yang-Mills-Higgs
equations only when the BPS condition is assumed. Let us approach
this in more detail.

Given the $SU(2)$ principal bundle on Minkowski space $M^{4}$, let
$A=A_{\mu}=A_{\mu}^{a}T^{a}$ be the connection on the bundle with
the field strength $F=F_{\mu\nu}=F_{\mu\nu}^{a}T^{a}$ and the standard
normalization of the generators of the gauge group is ${\rm Tr}(T^{a}T^{b})=\frac{1}{2}\delta_{ab}$.
For $SU(2)$, $T^{a}=\frac{\sigma^{a}}{2}$ The Lie algebra, as usual,
is defined via $[T^{a},T^{b}]=i\varepsilon_{abc}T^{c}$.

For the covariant derivative $D=D_{A}={\rm d}+A$ in this principal
$SU(2)$ bundle, the Yang-Mills equations read: \[
D_{A}F=0,\;\;\; D_{A}\star F=0\,\]
where as usual $\star$ is the Hodge $\star$-operator.

The Yang-Mills boson, as well the photon in the Maxwell equations,
has to be massless due to the gauge symmetry. To introduce a gauge
invariant mass terms into the theory one introduces the Higgs field
$\phi=\phi^{a}T^{a},a=1,2,3$ (summation), so that the action density
now becomes: \begin{equation}
L=-\frac{1}{2}{\rm Tr}F_{\mu\nu}F^{\mu\nu}+{\rm Tr}D_{\mu}\phi D^{\mu}\phi-V(\phi)=-\frac{1}{4}{\rm Tr}F_{\mu\nu}^{a}F^{a\mu\nu}+\frac{1}{2}{\rm Tr}(D^{\mu}\phi^{a})(D_{\mu}\phi^{a})-\tilde{V}(\phi)\,.\label{YM-0}\end{equation}
This is the Yang-Mills-Higgs action density. Suppose also that $\phi$
transforms in the adjoint of $SU(2)$, so that the potential can be
chosen as $V(\phi)=-\tilde{V}(\phi)=-\lambda(|\phi|^{2}-v^{2})^{2}$
and when $v^{2}\neq0$ the $SU(2)$ symmetry is broken to $U(1)$.
The action density now reads: \begin{equation}
L=(F,F)+(D\phi,D\phi)+V(\phi)=-\frac{1}{4}{\rm Tr}F_{\mu\nu}^{a}F^{a\mu\nu}+\frac{1}{2}{\rm Tr}(D^{\mu}\phi^{a})(D_{\mu}\phi^{a})-\lambda(|\phi|^{2}-v^{2})^{2}.\label{YM-00}\end{equation}
Now the mass term for the vector boson $A$, i.e. $m^{2}f(A^{2})$,
is introduced as follows. The minimal energy of the configuration
from (\ref{YM-0}) is attained for $V=F=D_{A}\phi=0$ and the Higgs
field \begin{equation}
\phi^{a}\underset{r\to\infty}{\longrightarrow}\frac{vr^{a}}{r}=\phi_{0}={\rm const}\,,\label{YM+2}\end{equation}
and in the vacuum background $\phi_{0}$ and $\phi$ are related by
gauge transformation, so that\[
D\phi_{0}={\rm d}\phi_{0}+A\phi_{0}=A\phi_{0}\,,\]
which means $(D\phi_{0},D\phi_{0})=(A\phi_{0},A\phi_{0})$. Introducing
it into the background action $L=(F,F)+(D\phi_{0},D\phi_{0})$ we
obtain the quadratic in $A$ mass term.

The vector bosons $A^{\pm}$ acquire charges due to the unbroken $U(1)$
which leaves the Higgs vacuum invariant (a direction in the isospin
space). So, one represents this subgroup by the rotations about the
direction in the isospace given by $\phi^{a}$ and the generator of
this rotation is $\frac{\phi^{a}T^{a}}{\sqrt{\phi^{a}\phi^{a}}}$
or $\frac{\phi^{a}T^{a}}{a}$ where $\phi^{a}\phi^{a}=a^{2}$. It
is the operator of the electric charge $Q$ at the same time. The
covariant derivative now reads: \[
D_{\mu}=\partial_{\mu}+iQA_{\mu}^{(e)}\]
where $A_{\mu}^{(e)}=\frac{1}{a}A_{\mu}^{a}\phi^{a}$ and $Q=e\frac{1}{a}\phi^{a}T^{a}$.
$A_{\mu}^{(e)}$ is the projection of $SU(2)$ $A_{\mu}^{a}$ onto
the direction of $\phi^{a}$. The corresponding strength is $F_{\mu\nu}^{a}=F_{\mu\nu}\frac{\phi^{a}}{a}$,
where \begin{equation}
F_{\mu\nu}^{a}=\partial_{\mu}A_{\nu}-\partial_{\nu}A_{\mu}+\frac{1}{a^{3}e}\varepsilon_{abc}\phi^{a}\partial_{\mu}\phi^{b}\partial_{\nu}\phi^{c}.\label{YM+1}\end{equation}
The non-trivial correction to the electrodynamic field strength $F_{\mu\nu}$
as seen in (\ref{YM+1}) vanishes in the topologically trivial sector
of the theory, while it is non-zero when the boundary condition are
as in (\ref{YM+2}). In this case the second pair of Maxwell equations
becomes $\partial^{\mu}\widetilde{F}_{\mu\nu}=k_{\nu}$ where \[
k_{\mu}=\frac{1}{2a^{3}e}\varepsilon_{\mu\nu\rho\sigma}\varepsilon_{abc}\partial^{\nu}\phi^{a}\partial^{\rho}\phi^{b}\partial^{\sigma}\phi^{c}\]
is the magnetic current absent in the pure electrodynamic case. Let
us observe that the current is expressible only via the Higgs field.
It is the first indication that, when the Dirac monopoles-magnetic
sources were found to be corresponding to some exotic $\mathbb{R}^{4}$s,
then we found the same mechanism for the Higgs field .

Next, turn to the monopole solutions. The EOM derived from (\ref{YM-0})
are: \begin{equation}
\begin{array}{c}
D_{A}F=0\\[5pt]
D_{A}\star F=-[\phi,D_{A}\phi]\\[5pt]
D_{A}\star D_{A}\phi=2\lambda\phi(|\phi|^{2}-v^{2})\end{array}\label{YM-1}\end{equation}
where the first equation is the Bianchi identity. The configuration
$(A,\phi)$ solving the equations (\ref{YM-1}) is the classical configuration
representing magnetic monopoles in the sense that every quantum theory
of magnetic monopoles should have the above configuration of fields
in the classical limit. $(A,\phi)$ is the configuration of fields
defining the magnetic monopole. The special static regular solutions
were numerically determined as: \[
\phi^{a}=\frac{r^{a}}{er^{2}}H(\xi),\;\; A_{n}^{a}=\varepsilon_{amn}\frac{r^{m}}{er^{2}}(1-K(\xi)),\;\; A^{a}=0,\;\xi=ver\,,\]
$H,K$ being some functions of $\xi$. These are the Polyakov-'t Hooft
solutions. Even though solving the YMH equations (\ref{YM-1}) in
general, on $H,K$, is intractable analytically, but the special limit
of Bogomol'nyi-Prasad-Sommerfield (BPS) can be solved. This case is
of  special importance for the connection with the exotic 4-geometries
on $\mathbb{R}^{4}$. Namely the BPS limit is determined by the vanishing
scalar potential $V(\phi)=0$ and the minimum of the energy is reached
by the static configurations which solves the 3-d Bogomol'nyi equation:
\begin{equation}
\varepsilon_{ijk}F_{ij}=\pm D_{k}\phi\,.\label{Bog-1}\end{equation}
These BPS monopoles are determined by the explicit $K$ and $H$,
namely: $K(\xi)=\frac{\xi}{\sinh{\xi}},\; H(\xi)=\xi\coth{\xi}-1\,.$

The crucial property of the BPS solutions is the connection between
mass and charge. The Bianchi identity and (\ref{Bog-1}) gives $D_{A_{n}}D_{A_{n}}\phi^{a}=0$,
so that the energy of the BPS monopole reads: $E=\frac{1}{2}\int{\rm d}x\partial_{n}\partial_{n}(\phi^{a}\phi^{a})=gv$.
Hence, when the dyon has electric $q$ and magnetic $g$ charges,
the mass of such BPS state is given by the Bogomol'nyi bound: \begin{equation}
M=v\sqrt{g^{2}+q^{2}}\,.\label{Bog-2}\end{equation}
The BPS solutions can be understood as the direct reduction of Euclidean
4-d YM theory into 3-dimensions. Namely, one considers pure YM equation
in the Euclidean $\mathbb{R}^{4}$ with an invariance of $x_{4}$
translation, so that the connection reads $A=A_{1}{\rm d}x_{1}+A_{2}{\rm d}x_{2}+A_{3}{\rm d}x_{3}+\phi{\rm d}x_{4}$.
$A_{i}$ and $\phi$ are now $SU(2)$-algebra valued functions on
$\mathbb{R}^{3}$ and the Euclidean action is written as $L=(F,F)+(D\phi,D\phi)$
where the curvature $F$ and the derivative $D$ are defined with
respect to the connection $A=A_{1}{\rm d}x_{1}+A_{2}{\rm d}x_{2}+A_{3}{\rm d}x_{3}$.
Now one sees that the YMH theory on $\mathbb{R}^{3}$ with $V(\phi)=0$
is the dimensional reduction of a pure YM on Euclidean $\mathbb{R}^{4}$.
Note that $V(\phi)=0$ is the BPS condition. Moreover, static monopole
BPS solutions are derived from the 4-d YM self-dual ones, i.e. $F=\star F$.
However, a more geometric picture is possible which is important in
this paper and is attained when classical solutions are seen as point-particles
in the space of solutions \cite{AH-1988}.

Given the space of the 3-d solutions $(A,\phi)$ modulo gauge invariance,
${\cal C}={\cal A}/{\cal G}$, the tangent vector is $(\overset{.}{A},\overset{.}{\phi})$
where one differentiate  with respect to some parameter of the path
in ${\cal C}$. Then, the (formal) Riemannian metric on ${\cal C}$:
\begin{equation}
|(\overset{.}{A},\overset{.}{\phi})|^{2}=\int_{\mathbb{R}^{3}}(\overset{.}{A},\overset{.}{A})+(\overset{.}{\phi},\overset{.}{\phi})\label{Lin}\end{equation}
and the potential energy $U=\frac{1}{2}\int_{\mathbb{R}^{3}}(F,F)+(D\phi,D\phi)$
are defined. We see that the motion of a particle on ${\cal C}$ with
the potential $U$ is governed by the solutions of the YMH equations
with the BPS condition. Thus, one determines the region (submanifold)
$M\subset{\cal C}$ on which $U$ attains the minimum. To ensure that
$U$ exists the following is assumed $|F|=O(r^{-2}),D\phi=O(r^{-2})$,
so that $|\phi|\underset{r\to\infty}{\longrightarrow}{\rm constant}=v^{2}$.
So, we are taking a 3-ball $B_{r}$ of radius $r$ and integrate \begin{equation}
\int_{B_{r}}(F,F)+(D\phi,D\phi)=\int_{B_{r}}(F-\star D\phi,F-\star D\phi)+2(\star D\phi,F)\label{YM-4}\end{equation}
where the second term is the surface integral over $S^{3}$, i.e.
$2\int_{S_{r}}(\phi,F)$, of the 2-form $(\phi,F)$ since the (total)
differential of it reads $d(\phi,F)=(D\phi,F)=\star(\star D\phi,F)$
and the Bianchi identity holds, i.e. $DF=0$. This is the topological
term and with its particular value $v^{2}=1$, i.e. $|\phi|\underset{r\to\infty}{\longrightarrow}1$,
one has: \[
\lim\limits _{r\to\infty}\int_{S_{r}}(\phi,F)=\pm4\pi k\]
where $\pm k$ can be seen as the Chern classes of the complex line
bundle on $S_{r}$ related with $\phi$ \cite{AH-1988}. This is the
magnetic charge of the solution. In this way the condition for the
absolute minimum of the potential energy $U$ is attained whenever
the first term in (\ref{YM-4}) vanishes, i.e. it holds: \begin{equation}
F=\star D\phi\label{YM-5}\end{equation}
which is the Bogomolnyi equation (\ref{Bog-1}).

Let us turn to the abelian $U(1)$ electrodynamic where the electric
field $E$ is the 1-form on $\mathbb{R}^{3}$, the magnetic field
$B$ is the 2-form and $A$ being the electromagnetic potential (2-form),
then, the field strength tensor, reads: $F=B+c{\rm d}t\wedge E$ and
$F={\rm d}A$. The pure Maxwell equations in the Minkowski spacetime
are: \[
{\rm d}F=0,\;\;{\rm d}\star F=0\,.\]
The Bogomolnyi equations for the abelian case become $B={\rm grad}\,{\phi}$,
and taking $\phi=\frac{k}{2r}$ one finds that this is the Dirac magnetic
monopole. Moreover, $B$ is the curvature of a connection on the line
bundle on $\mathbb{R}^{3}\setminus\{0\}$ which are classified by
$H^{3}(S^{3},\mathbb{Z})$. The appearance of this cohomology group
was the reason for the connection of the Dirac monopoles with the
special smooth $\mathbb{R}^{4}$ from the radial family, namely those
which are generated by the foliations with ${\rm GV}=k$ \cite{AsselmeyerKrol2009}.

The geometric perspective on the space of BPS monopoles solutions
is crucial for further exploring the relation of monopoles with exotic
4-geometries. We remark  that the geometry of the moduli space of
BPS monopoles can be exactly described and it is given by the complete
and non-singular metric. Such exact, precise and analytical results
are always of exceptional value and should be considered carefully.

Let us approach the geometry of the monopole moduli space in more
detail. For us the interesting case of $k=2$ static and weakly interacting
two BPS monopoles was described by Atiyah and Hitchin \cite{AH-1988}.
Following their analysis one observes that for $k=1$ the center of
the single BPS static monopole is fixed to a single point. So, the
reduced moduli space for $k=1$ is $M_{1}^{0}={\rm pt.}$, whereas,
taking into account the symmetries of this choice, the full moduli
space $M_{1}$ reads $M_{1}=\mathbb{R}^{3}\times S^{1}$. In general
the relation between the reduced and full moduli spaces can be written
as $M_{k}^{0}=\frac{M_{k}}{\mathbb{R}^{3}\times S^{1}}$. For us,
the interesting  case is the geometry of $M_{2}^{0}$. Analysis of
it is rather involved and based on profound works by Donaldson, Hurtubise
and Taubes among others (see, e.g. \cite{Taubes-1983,Donaldson-1984,Hurt-1985})
which were extensively used by Atiyah and Hitchin. As follows from
the work of Donaldson the parameter space of monopoles is very efficiently
modeled by the space of rational functions. Every $k$ BPS monopole,
as the solution of (\ref{YM-1}) with the BPS condition, determines
a connection, $A$, on the power of the line Hopf bundle $H$ on $S^{2}$,
i.e. $H^{k}\otimes H^{-k}$, which is prolonged radially to $S^{2}\times\mathbb{R}^{+}$
by the Higgs field $\phi$. Given $S^{2}\times\mathbb{R}^{+}=\mathbb{R}^{3}\setminus\{0\}$
we have the $SU(2)$ bundle on $S^{3}\setminus\{0\}$ with the $U(1)$
symmetry due to Higgs. If one takes the space of gauge equivalent
monopoles of charge $k$, $N_{k}$, then the above $M_{k}$ is fibered
over $N_{k}$ with fiber $S^{1}$. A point ${\rm pt.}$ in $M_{k}$
represents a monopole bundle $E_{k}$ up to some asymptotic isomorphism
of $E_{k}({\rm pt.})$ \cite{AH-1988}.

To relate this moduli space with rational functions it was proposed
to consider the scattering associated with the differential operator
$D_{A}$ defined along the line $u$ in $E_{k}$ and acting on the
sections of the bundle $E_{k}$ over $u$: \begin{equation}
D_{A}^{u}=\nabla_{A}^{u}-i\phi\label{Mon-1}\end{equation}
where $\nabla_{A}^{u}$ is the covariant derivation on $E_{k}$ along
$u$. Thus, given the magnetic BPS monopole $(A,\phi)$ as the solution
of the Bogomolny equation (\ref{Bog-1}) we can determine the differential
equation along $u$ as (\ref{Mon-1}). In fact one chooses the family
of parallel lines $u(t,z),t\in\mathbb{R},z\in\mathbb{C}$ in $\mathbb{R}^{3}$.
The equation $D_{A}^{u}s=0$ has two independent solutions $s_{0}(t,z),s_{1}(t,z)$
whose first solution decays at $t\to\infty$ so that $s_{0}(t,z)t^{-k/2}e^{t}\to e_{0}$.
Here $e_{0},e_{1}$ are the constant sections of the asymptotic gauge
for $t\to\infty$ where $e_{1}=e^{i\theta}e_{0}$ as usual. Next let
$s'_{0}(t,z)$ be a solution decaying at $t\to-\infty$ and $s'_{1}(t,z)$
a complementary solution, i.e. $s'_{0}(t,z)|t|^{-k/2}e^{|t|}\to e'_{0}$. 

Combining these solutions into a single scattering process one can
write $s'_{0}(t,z)=a(z)s_{0}(t,z)+b(z)s_{1}(t,z)$ where $b(z)$ is
a polynomial of degree $k$ which is the degree of the monopole bundle
$E_{k}$ and the charge of the monopole. Next one assigns the rational
function $S_{m}(z)=\frac{a(z)}{b(z)}\;{\rm mod}\, b(z)$ to the scattering
process driven by the monopole $m=(A,\phi)$. The remarkable theorem
of Donaldson holds true: \begin{theorem}[\cite{Donaldson-1984}] The
scattering functions $S_{m}(z),m\in M_{k}$ are rational functions
of degree $k$.

There exists a diffeomorphism $M_{k}\to R_{k}$ given by $m\to S_{m}(z)$
where $R_{k}$ is the space of all rational functions with $S_{m}(\infty)=0$.
\end{theorem} Thus, one can represent $M_{k}=R_{k}$ as the space
of rational functions in the normal form: \[
S(z)=\frac{\sum_{i=0}^{k-1}a_{i}z^{i}}{z^{k}+\sum_{i=o}^{k-1}b_{i}z^{i}}\]
providing the resultant $\Delta(a_{0},...,a_{k-1};\, b_{0},...,b_{k})$
vanishes. This means that $M_{k}$ is the open set in $\mathbb{C}^{2k}$.
In fact $M_{k}$ is the \emph{affine algebraic variety.} It is the
subvariety in $\mathbb{C}P^{2k}$ given by the complement of the subvariety
defined by homogenous equations $b_{k}\Delta=0$. The result by Yau
then says that $M_{k}$ has a Ricci-flat K\"ahler metric.

For $k=2$ we have \[
S(z)=\frac{a_{0}+a_{1}z}{z^{2}+b_{0}},\;\;\Delta=a_{0}^{2}+b_{0}a_{1}^{2}\]
so that  the quotient of the algebraic surface in $\mathbb{C}^{3}$:
$x^{2}-zy^{2}=1$ by the involution $(x,y,z)\to(-x,-y,z)$ gives the
manifold $M_{2}^{0}\subset\mathbb{C}P^{1}\times\mathbb{C}^{1}$ where
$(a_{0},a_{1})$ are homogenous coordinates in $\mathbb{C}P^{1}$.
The great achievement of Atiyah and Hitchin was the explicit description
of the metric on $M_{k}^{0}$. They showed that there exists a Riemannian
metric on $M_{k}^{0}$ which is finite, hyperk\"ahler, geodesically
complete and $SO(3)$-symmetric. However, it was the work of Takhtajan
\cite{Takhtajan-1992} showing that this metric is precisely  the
AH gravitational instanton given in terms of quasi-modular forms.

To show finiteness of the metric one can turn to the zero-modes generating
metric as in (\ref{Lin}), and investigate their $L^{2}$-norms. This
is not an easy task since the equations are defined on the non-compact
$\mathbb{R}^{3}$ and the gauge freedom should be included. However,
Taubes performed the analysis successfully showing the finiteness
of the metric of $M_{k}^{0}$ \cite{Taubes-1983}.

To find a suitable description of the tangent space modulo gauges,
one can turn to the linearization of the Bogomolny equations. For
a BPS monopole $m=(A,\phi)$ the tangent space for $M_{k}$ in $m$
is the space $T_{m}$ of all pairs $(a,\psi)$ which fulfills the
linearized equation \cite{AH-1988,Stuart-1994}: \[
\star D_{A}a-D_{A}\psi+[\phi,a]=0\,.\]
This should be augmented by the equation \[
\star D_{A}\star a+[\phi,\psi]=0\]
which expresses the fact that $(a,\psi)$ is orthogonal to the directions
generated by the infinitesimal elements in gauge algebra. Here $a$
is a Lie-algebra-valued 1-form and $\psi$ is a function also with
values in the Lie algebra of $SU(2)$. Taubes showed that the dimension
of $T_{m}$ is $4k$. For $I,J,K$ (the base in  the quaternion space
${\textbf{H}}$) one can write the 1-form $a=\alpha{\rm d}x+\alpha{\rm d}y+\alpha{\rm d}z$
and $\psi$, in one expression: $\psi+\alpha I+\beta J+\gamma K$.
In this way $(a,\psi)$ becomes a function $(a,\psi):\mathbb{R}^{3}\to{\rm su}(2)\times{\textbf{H}}$.
Both equations are invariant with respect to the action of $I,J,K$,
i.e. $T_{m}$ is the vector space over ${\textbf{H}}$. Now one introduces
 the natural $L^{2}$-metric on this vector space and $I,J,K$ act
as isometries. Given the norm on $M_{k}$ defined as $||(a,\psi)||_{m}^{2}=||D_{A}a||_{L^{2}}^{2}+||D_{A}\psi||_{L^{2}}^{2}+||[\phi,a]||_{L^{2}}^{2}+||[\phi,\psi]||_{L^{2}}^{2}$
one shows that the above metric is complete \cite{Taubes-1983}.

$I^{2}=J^{2}=K^{2}=-1$ so there are 3 almost complex structures which
are actually  integrable (Donaldson) giving rise to 3 complex structures
on $M_{k}$ and to three K\"ahler forms as well. This means that $M_{k}$
and $M_{k}^{0}$ are hyperk\"ahler complex manifolds. It is best seen
from the infinite-dimensional hyperk\"ahler quotient construction where
the 3-components of the Bogomolny equation give rise to the 3-moment
maps with values in the dual of the gauge algebra \cite{AH-1988}.
We remark that the  hyperk\"ahler $M_{k}$ is Ricci-flat.

$SO(3)$ acts isometrically on $M_{k}$. This action is given by 
rotating the complex structures $I,J,K$ on $M_{k}$. One projects
$M_{k}$ on $\mathbb{R}^{3}$ since the universal cover $\tilde{M_{k}^{0}}$
acts trivially on $\mathbb{R}^{3}$ so, the projection is the assignment
of the center to a monopole in a $SO(3)$-equivariant way \cite{AH-1988}.
We have the metric on $M_{2}^{0}$ which has $SO(3)$ as a group of
its isometries. Then we are looking for the solutions of anti-self-dual
Einstein equations, which are $SO(3)$-invariant. The differential
equations of this metric in 4-d are reduced to a system of ordinary
differential equations. It follows from \cite{GP-1979} that the metric
written in terms of Maurer-Cartan basis of $SU(2)$ (\ref{o-4}),
reads: \begin{equation}
{\rm d}s^{2}={\rm d}t^{2}+a(\sigma^{1})^{2}+b(\sigma^{2})^{2}+c(\sigma^{3})^{2}\,.\label{0-55}\end{equation}
Rewriting it in terms of $\Theta^{i},i=1,2,3$, such that ${\rm d}t'=\sqrt{\Theta^{1}\Theta^{2}\Theta^{3}}{\rm d}t$,
one arrives at the ordinary differential equations (\ref{o-11}) and
the case $\Theta^{1}\neq\Theta^{2}\neq\Theta^{3}$ is crucial (the
$SO(3)$-orbits are 3-dimensional). That is, the AH geometry of $M_{2}^{0}$
is precisely that of ${\cal M}_{3}\times\mathbb{R}$ and the metric
is written in terms of quasi-modular series $E_{2}$. Thus we close
the circle of the argumentation in the paper and arrive at quasi-modularity
and the geometry we started with.

Finally let us comment on the BPS condition as allowing for the direct
connection of magnetic (charged) monopoles with exotic 4-geometry.
Namely, the Lagrangian (\ref{YM-00}) gives rise to the energy-momentum
tensor: \begin{equation}
T_{\mu\nu}=F_{\mu\rho}^{a}F_{\nu}^{a\rho}+D_{\mu}\phi^{a}D_{\nu}\phi^{a}-g_{\mu\nu}{L}\label{EMT-1}\end{equation}
so that the static energy in terms of $E^{ai}=F^{ai0}$ and $B^{ai}=-\frac{1}{2}\varepsilon^{ijk}F_{ajk}$
fields reads: \[
E=\int{\rm d}^{3}xT_{00}=\int{\rm d}^{3}x\frac{1}{2}(E_{i}^{a}E_{i}^{a}+B_{i}^{a}B_{i}^{a}+(D_{i}\phi^{a})(D_{i}\phi^{a}))+\frac{\lambda}{4}(\phi^{a}\phi^{a}-v^{2})^{2}\,.\]

For simplicity and due to the absence of explicit formulas on an exotic
$\mathbb{R}^{4}$, we assume additionally that this $T_{\mu\nu}$
is precisely generated by the change of the exotic smoothness of $e$
into the standard $\mathbb{R}^{4}$ as discussed before. This happens
in the FTL limit of GR. Thus, we can consider both, the 'breaking'
of the exotic 4-smooth structure which results in the standard one,
and the appearing of the Higgs axis with non-zero vev breaking the
$SU(2)$ symmetry, as very close related. The appearance of the Higgs
direction seen in exotic 4-d, automatically reduces the smoothness
into the standard one. Thus, $T_{\mu\nu}$ in the theory with a Higgs
is generated by $T_{\mu\nu}$ from an exotic $\mathbb{R}^{4}$. However,
the BPS bound (\ref{Bog-2}) gives also charges from masses in the
static limit. This purely formal relation indicates  the possibility
to generate monopoles as charged configurations from exotic 4-geometry.
We will discuss this physically intriguing point further in the next
section.

The extension of the geometry of $M_{2}^{0}$ over a regime of quantum
dynamics of interacting monopoles is possible and was indeed performed
\cite{Manton-1986,Schroer-1991,Stuart-1994}. This extension shows
the true magic of the geometry of $M_{2}^{0}$ which is based on the
geodesic approximation to interacting almost static quantum monopoles,
the program first proposed by Manton. In particular, the role of the
geometry of $M_{2}^{0}$ in the quantum regime is reflected in considering
the Schr{\"o}dinger operator as being proportional to the covariant
Laplacian on $M_{2}^{0}$. Let $\xi_{1},\xi_{2},\xi_{3}$ be the dual
vector fields to the Maurer-Cartan 1-forms $\sigma_{1},\sigma_{2},\sigma_{3}$
as in (\ref{o-4}) for $SU(2)$, and $f=\frac{-\Theta^{2}}{r}$ Then
the Schr{\"o}dinger equation reads in the coordinates introduced
in Sec. \ref{Gr-I}: \[
-\frac{1}{\Theta^{1}\Theta^{2}\Theta^{3}f}\frac{\partial}{\partial r}\left(\frac{\Theta^{1}\Theta^{2}\Theta^{3}}{f}\frac{\partial\Psi}{\partial r}\right)-\left(\frac{\xi_{1}^{2}}{(\Theta^{1})^{2}}+\frac{\xi_{2}^{2}}{(\Theta^{2})^{2}}+\frac{\xi_{3}^{2}}{(\Theta^{3})^{2}}\right)\Psi=\pi\frac{E}{\hbar^{2}}\Psi  \,.\]
In certain approximations one obtains the Taub-Nut geometry of quantum
interacting monopoles \cite{Manton-1986}. In fact, this quantum case
can be solved exactly. However, it is a merely approximation for the
complete quantum dynamics of monopoles in AH metric of $M_{2}^{0}$,
in which case, quantum geodesics show also a chaotic behavior. This
last fact can be (possibly) understood in terms of wild embeddings
characterizing the exotic 4-geometry. This interesting point will
be approached in a separate publication. Moreover, when dealing with
quantum aspects of the geometry of moduli spaces there appears a mechanism
where angular momentum generates electric charges of the monopoles/dyons
configurations \cite{Schroer-1991}, which is the case for the energy
generating the charges discussed above.

On the other hand, due to the Conjecture \ref{conj} 1 one can approach
the regime of quantum gravity via magnetic monopoles. The meaning
of this will be discussed elsewhere, however some remarks are placed
also in the next section.

\section{Discussion}

We obtained a rather remarkable way of introducing the Higgs field
into the $SU(2)$ YM theory on 4-d exotic $e$ when changing the smoothness
into the standard one. $SU(2)$ magnetic monopoles of  topological
charge $k$ determine the geometry of their (reduced) moduli spaces
$M_{k}^{0}$. This metric is complete, hyperk\"ahler and defined on
the whole $M_{k}^{0}$ and for $k=2$ matches precisely the geometry
of the gravitational instanton emerging from exotic $\mathbb{R}^{4}$.
The appearance of this geometry determines the dynamics of low energy
magnetic monopoles and is the strong indication for magnetic monopoles
as generating the geometry. The metric of it governs the interactions
of low energy $k=2$ magnetic monopoles. Any quantum theory of magnetic
$k=2$ monopoles has to give this geometry in the classical limit
\cite{AH-1988}. But this geometry governs also the quantum dynamics
of monopoles \cite{Schroer-1991}. In our case this geometry is generated
by the 4-d exotic $e$ and by the dynamics of GR in the FTL limit.
This means that the exotic $e$ captures the limit of the dual magnetic
dynamics of point-like monopoles and this can be uniquely prolonged
into the regime where monopoles are not point-particles and are rather
quantum entities \cite{Schroer-1991}. Given the 4-geometric origins
of these monopoles, the Higgs field appears from the classical geometry
on $e$, too. We expect that this can be somehow extended over quantum
regimes of the theories: QG and standard model of particles and fields.
One indication of this expectation is the relation between magnetic
charges of Dirac monopoles and the GV classes of the foliations assigned
to the exotic $\mathbb{R}^{4}$s from the radial family of deMichelis-Freedman
type \cite{AsselmeyerKrol2009}. Another hints are derivable from
the relation of these exotic $\mathbb{R}^{4}$s with superstring theory.
Again foliations and wild embeddings are the crucial geometrical base
for the correspondence \cite{AsselmeyerKrol2011,AsselmKrol2011f,AsselmeyerKrol2011b}.
In this paper we showed how the semiclassical approach to QG is well-suited
for analyzing the role of the geometry of $e$ in FTL of GR.

The arguments of this paper indicate the following possibility: starting
with the pure $SU(2)$ YM on the exotic $\mathbb{R}^{4}$ then changes
the smoothness into the standard $\mathbb{R}^{4}$. This can switch
on the Higgs field and turn the theory to Yang-Mills-Higgs theory.
It breaks exotic smoothness and massive vector bosons appear. The
Higgs field breaks the $SU(2)$ symmetry of YM theory on $e$, but
also $\phi$ breaks the exotic structure of $e$ enforcing the reduction
of the $SU(2)$ symmetry to $U(1)$ symmetry along a line in the isospace.
The appearance of the Polyakov-'t Hooft magnetic monopoles driven by
the 4-geometry can be further discussed from many different physical
perspectives, leading to some unifying power.

\emph{Cosmology and GUT theories.} YMH theory is the modification
of the earlier attempt of Georgi and Glashow (1972) to incorporate
the Higgs mechanism in the YM theory with $SO(3)$ gauge group. The
Polyakov-'t Hooft modification relies on this attempt but with the
$SU(2)$ gauge group. This theory does not match the experimental
predictions correctly, since the mass of monopoles from YMH would
be in the testable range but are not found there. Furthermore the
gauge group of SM is not just $SU(2)$. The $SU(5)$ GUT theory (1974)
of Georgi and Glashow also predicts the existence of magnetic monopoles
of this kind. As is known, the $SU(5)$ GUT predicts instability of
proton having a finite live-time which was also falsified experimentally.
However, every GUT-type theory with a larger group than $SU(2)$ 
necessarily predicts the existence of magnetic monopoles when the
symmetry is broken and the mass of them can be much higher. Larger
groups than $SU(5)$ groups in GUT theories draw this limit and the
live-time of proton behind the testable border . The original Dirac
monopoles (1931) introduced on the base of quantum mechanics does
not have assigned strictly defined masses.

The absence of magnetic monopoles in the universe enforced researchers
to find out a reason explaining the low density of monopoles, provided
they exist. This led to the theory of inflation, which is the cornerstone
of modern cosmology. Here, the possibility appears that monopoles
are replaced by 4-geometry which is not limited only to the early
universe epoch. So far, there are theoretical arguments that 4-geometry
is assigned to condensed effective matter, and this geometry would
have physical meaning. Moreover, it is possible that inflation is
derived from nonstandard smooth structures on some 4-manifolds. It
was  described for the case of fake Freedman 'ends' $S_{\theta}^{3}\times\mathbb{R}$
\cite{AsselmKrol-2012e}. These topics are currently under active
development.

Gravity confined to the exotic $e$, YM and Higgs fields related to
magnetic monopoles and to $e$, all these constituents suggest to
consider the theory unifying them, i.e. Einstein-Yang-Mills-Higgs
model (EYMH) (see e.g., \cite{Dehnen-2008}). This theory introduces
essential coupling of Higgs to curvature and is used for approaching
 some important questions in cosmology and for modeling gravity in
condensed matter phenomena (see e.g., \cite{Dehnen-2008}). The contributions
of the Higgs fields to the total stress-energy tensor of the system
are considered in many models as dominating at the inflation stage.
The models with Higgs fields appeared in cosmology in the theory of
dark matter and dark energy but the discovery of accelerated expansion
of the present Universe showed that non-minimal Einstein-Yang-Mills-Higgs
models are important in the search for an explanation of the dark
energy phenomenon. One derives exact solutions for metrics respecting
Higgs background which might be useful in search for metrics on exotic
$\mathbb{R}^{4}$s. Moreover, cosmological solutions with naked singularities
appear in this framework and naked singularities must also be present
in GR on exotic $\mathbb{R}^{4}$ \cite{Asselm-Krol-2011}. These
new attractive approach to exotic 4-spaces in the above context of
cosmology will be addressed in a separate publication.

\emph{Condensed matter, spin-ice and cold atoms.} So far, the search
for magnetic monopoles as fundamental particles failed. That is why
people started with attempts to realize monopoles  as emergent phenomenon,
i.e. as manifestations of the correlations present in a strongly interacting
many-body system. That is why magnetic monopoles-like states were
expected to be found, and are indeed reported as appearing, among
states realized on the, so called, spin-ice systems \cite{Castelnovo-2008,Jaubert-2009,Giblin-2011}.
This seems to be a very promising approach which is important also
from the exotic 4-geometry point of view. Moreover, one faces growing
recent activity on experimental approaching the degenerate quantum
gases as revealing the potential of the cold atom systems. They show
quite universal features and serve as quantum simulators for ideas
far beyond the usual condensed matter phenomena. In particular, non-Abelian
gauge potentials could be realized in the effective description of
atoms with degenerate internal degrees of freedom coupled to spatially
varying laser fields. Namely, dilute Bose-Einstein condensates (BECs)
of alkali atoms (with a hyperfine spin degree of freedom) combine
magnetic and superfluid order. The order parameter describing such
systems is invariant under global symmetries of a non-Abelian group
and thus, monopoles can occur if this symmetry is broken. Thus, artificially
generated gauge fields in spinor BECs, can provide an alternative
method to realize a magnetic monopole and in the simplest case of
spin-1 condensate, a variety of different topological defects indeed
are expected. These are global monopoles, non-Abelian magnetic monopoles,
global textures and analogies to Dirac monopoles \cite{Kawa-2008,Piet-2009,Savage-2003}.
An experimental realization of any of these topological states still
remains a challenge in the field of cold atoms, though, from the point
of view of ideas developed in this paper, the appearance of Dirac
and non-abelian 't Hooft-Polyakov magnetic monopoles in the above condensates
indicates  a new and fundamental connection between 3-d physics and
4-d geometry. The link is given by exotic 4-geometry and the approach
deals with the collective states of effective matter. Certainly, the
fundamental connection between 4-d and 3-d physics was predicted already
by Einstein's relativity theories. However, here the connection is
prolonged somehow over the low energy, static regimes, at least for
the magnetic monopoles. Previously, based on the algebraic considerations,
it was also proposed to relate 4-geometries of certain exotic $\mathbb{R}^{4}$s
with the Kondo states created by effective electrons in the multichannel
Kondo effect \cite{AsselmKrol11}. Notion of fundamental microscopic
matter is modified, as is gravity connected with such matter. Even
static states of 3-d magnetic matter can be rooted in 4-d non-flat
geometries.

\emph{The standard model of particles and QG.} Magnetic monopoles
considered as fundamental point-particles, generates the 4-geometry
which has also gravitational meaning seen via the geometry and curvature
of exotic $\mathbb{R}^{4}$ (cf. \cite{AsselmKrol11}). Such situation
indicates on the possible two regimes of gravity, one connected with
'microscopic' matter of the standard model and with the standard smoothness
on $\mathbb{R}^{4}$, and the other, connected with the effective
matter and exotic $\mathbb{R}^{4}$s. Because of that one could consider
two different smoothness structures on which two theories, gravity
and SM of particles, are built. This difference is realized in the
smoothness assigned to GR and SM. From this point of view one can
approach  the difficulties of a theory which would unify QFT and GR.
Some new ingredients resulting from the approach above might be crucial
for unifying the theories (cf. \cite{QG-2012}).

\emph{Supersymmetry and SW theory.} The FTL was recently considered
in the context of supersymmetric ${\cal N}=2$ YM theories on $e$
and, in fact, this was the motivation for the present work. Gravitational
corrections to Seiberg-Witten theory on $\mathbb{R}^{4}$ were determined
as the corrections derived from FTL limit of SYM on $e$ \cite{AsselmKrol2012b,AsselmKrol2012c}.
The supersymmetry played a crucial role in the derivation. It is known
that SW theory exhibits the confinement and the mass gap but, again,
supersymmetry is crucial. One could wonder weather the direct connection
of exotic $e$ with magnetic monopoles and the possibility to formulate
YM theory on $e$ are sufficient for the generation of confinement
mechanism without supersymmetry. The condensation of monopoles would
be replaced by 4-geometric structures. If such attempt results in
a theory bearing realistic features would be a new insight into the
confinement problem. This requires, however, a careful analysis and
will be discussed elsewhere. Despite the substantial theoretical effort
it would be very desirable and interesting to indicate and analyze
purely experimental signals in favor to the role of 4-geometry in
various branches of physics, from low energy, particle physics to
global gravitational effects. We hope that at near future we will
exhibit some of them.

Our knowledge of 3-d geometries became rather complete mainly due
to the seminal works by Thurston and Perelman who proved the geometrization
conjecture of Thurston. The 4-d case is still mysterious and there
remains many unanswered questions especially for open 4-manifolds.
The results in this paper as well in some related papers show that
certain 4-geometries on $\mathbb{R}^{4}$ are able to code dynamics
of $3+1$ theories with many, also non-perturbative, aspects. These
geometries are also assigned to low energy states of condensed matter
and may play an active role in redefining fundamental matter and gravity
at this regime. This is just the beginning of uncovering  the meaning
of this code. It seems that much more can be understood in the near
future by exploring these questions.
\acknowledgements{JK thanks Sebastian Zaj\k{a}c for very fruitful discussion of some topics in the paper}


\begin{thebibliography}{43}
\expandafter\ifx\csname natexlab\endcsname\relax\def\natexlab#1{#1}\fi
\expandafter\ifx\csname bibnamefont\endcsname\relax
  \def\bibnamefont#1{#1}\fi
\expandafter\ifx\csname bibfnamefont\endcsname\relax
  \def\bibfnamefont#1{#1}\fi
\expandafter\ifx\csname citenamefont\endcsname\relax
  \def\citenamefont#1{#1}\fi
\expandafter\ifx\csname url\endcsname\relax
  \def\url#1{\texttt{#1}}\fi
\expandafter\ifx\csname urlprefix\endcsname\relax\def\urlprefix{URL }\fi
\providecommand{\bibinfo}[2]{#2}
\providecommand{\eprint}[2][]{\url{#2}}

\bibitem[{\citenamefont{Nekrasov}(2004)}]{Nekr2002e}
\bibinfo{author}{\bibfnamefont{N.~A.} \bibnamefont{Nekrasov}},
  \bibinfo{journal}{Adv. Theor. Math. Phys.} \textbf{\bibinfo{volume}{7}},
  \bibinfo{pages}{831} (\bibinfo{year}{2004}),
  \bibinfo{note}{arXiv:hep-th/0206161}.

\bibitem[{\citenamefont{Eguchi et~al.}(1980)\citenamefont{Eguchi, Gilkey, and
  Hanson}}]{EGH-1980}
\bibinfo{author}{\bibfnamefont{T.}~\bibnamefont{Eguchi}},
  \bibinfo{author}{\bibfnamefont{P.~B.} \bibnamefont{Gilkey}},
  \bibnamefont{and} \bibinfo{author}{\bibfnamefont{A.~J.}
  \bibnamefont{Hanson}}, \bibinfo{journal}{Phys. Rep.}
  \textbf{\bibinfo{volume}{66}}, \bibinfo{pages}{213} (\bibinfo{year}{1980}).

\bibitem[{\citenamefont{Hawking}(1977)}]{Hawking-1977}
\bibinfo{author}{\bibfnamefont{S.~W.} \bibnamefont{Hawking}},
  \bibinfo{journal}{Phys. Lett.} \textbf{\bibinfo{volume}{60A}},
  \bibinfo{pages}{81} (\bibinfo{year}{1977}).

\bibitem[{\citenamefont{Gibbons and Pope}(1979)}]{GP-1979}
\bibinfo{author}{\bibfnamefont{G.~W.} \bibnamefont{Gibbons}} \bibnamefont{and}
  \bibinfo{author}{\bibfnamefont{C.~N.} \bibnamefont{Pope}},
  \bibinfo{journal}{Commun. Math. Phys.} \textbf{\bibinfo{volume}{66}},
  \bibinfo{pages}{267} (\bibinfo{year}{1979}).

\bibitem[{\citenamefont{Witten}(1985)}]{Witten85}
\bibinfo{author}{\bibfnamefont{E.}~\bibnamefont{Witten}},
  \bibinfo{journal}{Commun. Math. Phys.} \textbf{\bibinfo{volume}{{\bf 100}}},
  \bibinfo{pages}{197} (\bibinfo{year}{1985}).

\bibitem[{\citenamefont{Milnor}(1962)}]{Milnor1962}
\bibinfo{author}{\bibfnamefont{J.}~\bibnamefont{Milnor}}, \bibinfo{journal}{Am.
  J. Math.} \textbf{\bibinfo{volume}{84}}, \bibinfo{pages}{1}
  (\bibinfo{year}{1962}).

\bibitem[{\citenamefont{DeMichelis and Freedman}(1992)}]{DeMichFreedman1992}
\bibinfo{author}{\bibfnamefont{S.}~\bibnamefont{DeMichelis}} \bibnamefont{and}
  \bibinfo{author}{\bibfnamefont{M.}~\bibnamefont{Freedman}},
  \bibinfo{journal}{J. Diff. Geom.} \textbf{\bibinfo{volume}{35}},
  \bibinfo{pages}{219} (\bibinfo{year}{1992}).

\bibitem[{\citenamefont{Asselmeyer-Maluga and Kr{\'o}l}(2010)}]{AssKrol2010ICM}
\bibinfo{author}{\bibfnamefont{T.}~\bibnamefont{Asselmeyer-Maluga}}
  \bibnamefont{and} \bibinfo{author}{\bibfnamefont{J.}~\bibnamefont{Kr{\'o}l}},
  in \emph{\bibinfo{booktitle}{International Congress of Mathematicians ICM
  2010, Hyderabad, India, Short Communications Abstracts Book}}, edited by
  \bibinfo{editor}{\bibfnamefont{E.~R.} \bibnamefont{Bathia}}
  (\bibinfo{organization}{Hindustan Book Agency}, \bibinfo{year}{2010}), p.
  \bibinfo{pages}{400}.

\bibitem[{\citenamefont{Asselmeyer-Maluga and
  Kr{\'o}l}(2012)}]{AsselmKrol2011d}
\bibinfo{author}{\bibfnamefont{T.}~\bibnamefont{Asselmeyer-Maluga}}
  \bibnamefont{and} \bibinfo{author}{\bibfnamefont{J.}~\bibnamefont{Kr{\'o}l}},
  \bibinfo{journal}{Int. J. Geom. Meth. Mod. Phys.}
  \textbf{\bibinfo{volume}{9}} (\bibinfo{year}{2012}),
  \bibinfo{note}{arXiv:1102.3274}.

\bibitem[{\citenamefont{Asselmeyer-Maluga and
  Kr{\'o}l}(2011{\natexlab{a}})}]{AsselmeyerKrol2011}
\bibinfo{author}{\bibfnamefont{T.}~\bibnamefont{Asselmeyer-Maluga}}
  \bibnamefont{and} \bibinfo{author}{\bibfnamefont{J.}~\bibnamefont{Kr{\'o}l}},
  \bibinfo{journal}{Int. J. Mod. Phys. A} \textbf{\bibinfo{volume}{26}},
  \bibinfo{pages}{1375} (\bibinfo{year}{2011}{\natexlab{a}}),
  \bibinfo{note}{arXiv:1101.3169}.

\bibitem[{\citenamefont{Asselmeyer-Maluga and
  Kr{\'o}l}(2011{\natexlab{b}})}]{AsselmeyerKrol2011b}
\bibinfo{author}{\bibfnamefont{T.}~\bibnamefont{Asselmeyer-Maluga}}
  \bibnamefont{and} \bibinfo{author}{\bibfnamefont{J.}~\bibnamefont{Kr{\'o}l}},
  \bibinfo{journal}{Int. J. Mod. Phys. A} \textbf{\bibinfo{volume}{26}},
  \bibinfo{pages}{3421} (\bibinfo{year}{2011}{\natexlab{b}}),
  \bibinfo{note}{arXiv:1105.1557}.

\bibitem[{\citenamefont{Asselmeyer-Maluga
  et~al.}(2013)\citenamefont{Asselmeyer-Maluga, Gusin, and
  Kr\'ol}}]{AsselmKrol2011f}
\bibinfo{author}{\bibfnamefont{T.}~\bibnamefont{Asselmeyer-Maluga}},
  \bibinfo{author}{\bibfnamefont{P.}~\bibnamefont{Gusin}}, \bibnamefont{and}
  \bibinfo{author}{\bibfnamefont{J.}~\bibnamefont{Kr\'ol}},
  \bibinfo{journal}{will appear in Int. J. Geom. Meth. Mod. Phys.}
  \textbf{\bibinfo{volume}{10 No 1}} (\bibinfo{year}{2013}),
  \bibinfo{note}{arXiv: 1109.1973}.

\bibitem[{\citenamefont{Asselmeyer-Maluga and
  Kr\'ol}(2012{\natexlab{a}})}]{AsselmKrol2012b}
\bibinfo{author}{\bibfnamefont{T.}~\bibnamefont{Asselmeyer-Maluga}}
  \bibnamefont{and} \bibinfo{author}{\bibfnamefont{J.}~\bibnamefont{Kr\'ol}}
  (\bibinfo{year}{2012}{\natexlab{a}}), \bibinfo{note}{{T}owards superconformal
  and quasi-modular representation of exotic smooth $\mathbb{R}^4$ from
  superstring theory I, arXiv:1207.4603}.

\bibitem[{\citenamefont{Asselmeyer-Maluga and
  Kr{\'o}l}(2009)}]{AsselmeyerKrol2009}
\bibinfo{author}{\bibfnamefont{T.}~\bibnamefont{Asselmeyer-Maluga}}
  \bibnamefont{and} \bibinfo{author}{\bibfnamefont{J.}~\bibnamefont{Kr{\'o}l}}
  (\bibinfo{year}{2009}), \bibinfo{note}{arXiv: 0904.1276}.

\bibitem[{\citenamefont{Asselmeyer-Maluga and
  Kr\'ol}(2012{\natexlab{b}})}]{AsselmKrol2012c}
\bibinfo{author}{\bibfnamefont{T.}~\bibnamefont{Asselmeyer-Maluga}}
  \bibnamefont{and} \bibinfo{author}{\bibfnamefont{J.}~\bibnamefont{Kr\'ol}}
  (\bibinfo{year}{2012}{\natexlab{b}}), \bibinfo{note}{{T}owards superconformal
  and quasi-modular representation of exotic smooth $\mathbb{R}^4$ from
  superstring theory II, arXiv:1207.4603}.

\bibitem[{\citenamefont{Zagier}(2008)}]{Zagier2008}
\bibinfo{author}{\bibfnamefont{D.}~\bibnamefont{Zagier}},
  \emph{\bibinfo{title}{The 1-2-3 of Modular Forms}}
  (\bibinfo{publisher}{Springer}, \bibinfo{year}{2008}), chap.
  \bibinfo{chapter}{Elliptic modular forms and their applications}.

\bibitem[{\citenamefont{Atiyah and Hitchin}(1985)}]{AH-1985}
\bibinfo{author}{\bibfnamefont{M.~F.} \bibnamefont{Atiyah}} \bibnamefont{and}
  \bibinfo{author}{\bibfnamefont{N.~J.} \bibnamefont{Hitchin}},
  \bibinfo{journal}{Phys. Lett.} \textbf{\bibinfo{volume}{107A}},
  \bibinfo{pages}{21} (\bibinfo{year}{1985}).

\bibitem[{\citenamefont{Atiyah and Hitchin}(1988)}]{AH-1988}
\bibinfo{author}{\bibfnamefont{M.}~\bibnamefont{Atiyah}} \bibnamefont{and}
  \bibinfo{author}{\bibfnamefont{N.}~\bibnamefont{Hitchin}},
  \emph{\bibinfo{title}{The geometry and dynamics of magnetic monopoles}}
  (\bibinfo{publisher}{M. B. Porter Lectures, Princeton, University Press,
  Princeton, NJ}, \bibinfo{year}{1988}).

\bibitem[{\citenamefont{Taubes}(1984)}]{Taubes-1984}
\bibinfo{author}{\bibfnamefont{C.~H.} \bibnamefont{Taubes}},
  \bibinfo{journal}{J. Differential Geom.} \textbf{\bibinfo{volume}{19}},
  \bibinfo{pages}{517} (\bibinfo{year}{1984}).

\bibitem[{\citenamefont{Tsukamoto}(2010)}]{Tsukamoto2010}
\bibinfo{author}{\bibfnamefont{M.}~\bibnamefont{Tsukamoto}}
  (\bibinfo{year}{2010}), \bibinfo{note}{arxiv:1004.3394}.

\bibitem[{\citenamefont{Oh et~al.}(2011)\citenamefont{Oh, Park, and
  Yang}}]{Oh-2011}
\bibinfo{author}{\bibfnamefont{J.~J.} \bibnamefont{Oh}},
  \bibinfo{author}{\bibfnamefont{C.}~\bibnamefont{Park}}, \bibnamefont{and}
  \bibinfo{author}{\bibfnamefont{H.~S.} \bibnamefont{Yang}},
  \bibinfo{journal}{JHEP} \textbf{\bibinfo{volume}{1104}}, \bibinfo{pages}{087}
  (\bibinfo{year}{2011}), \bibinfo{note}{arXiv:1101.1357}.

\bibitem[{\citenamefont{Asselmeyer-Maluga and Kr\'ol}(2011)}]{Asselm-Krol-2011}
\bibinfo{author}{\bibfnamefont{T.}~\bibnamefont{Asselmeyer-Maluga}}
  \bibnamefont{and} \bibinfo{author}{\bibfnamefont{J.}~\bibnamefont{Kr\'ol}}
  (\bibinfo{year}{2011}), \bibinfo{note}{arXive:1112.4882}.

\bibitem[{\citenamefont{Brans}(1994)}]{Bra:94b}
\bibinfo{author}{\bibfnamefont{C.}~\bibnamefont{Brans}}, \bibinfo{journal}{J.
  Math. Phys.} \textbf{\bibinfo{volume}{{\bf 35}}}, \bibinfo{pages}{5494}
  (\bibinfo{year}{1994}).

\bibitem[{\citenamefont{Asselmeyer}(1996)}]{Ass:96}
\bibinfo{author}{\bibfnamefont{T.}~\bibnamefont{Asselmeyer}},
  \bibinfo{journal}{Class. Quant. Grav.} \textbf{\bibinfo{volume}{{\bf 14}}},
  \bibinfo{pages}{749 } (\bibinfo{year}{1996}).

\bibitem[{\citenamefont{S{\l}adkowski}(2001)}]{Sladkowski2001}
\bibinfo{author}{\bibfnamefont{J.}~\bibnamefont{S{\l}adkowski}},
  \bibinfo{journal}{Int.J. Mod. Phys. D} \textbf{\bibinfo{volume}{10}},
  \bibinfo{pages}{311} (\bibinfo{year}{2001}).

\bibitem[{\citenamefont{Asselmeyer-Maluga and Brans}(2011)}]{AsselBrans2011}
\bibinfo{author}{\bibfnamefont{T.}~\bibnamefont{Asselmeyer-Maluga}}
  \bibnamefont{and} \bibinfo{author}{\bibfnamefont{C.~H.} \bibnamefont{Brans}}
  (\bibinfo{year}{2011}), \bibinfo{note}{arXiv:1101.3168}.

\bibitem[{\citenamefont{Taubes}(1983)}]{Taubes-1983}
\bibinfo{author}{\bibfnamefont{C.~H.} \bibnamefont{Taubes}},
  \bibinfo{journal}{Commun. Math. Phys.} \textbf{\bibinfo{volume}{91}},
  \bibinfo{pages}{235} (\bibinfo{year}{1983}).

\bibitem[{\citenamefont{Donaldson}(1984)}]{Donaldson-1984}
\bibinfo{author}{\bibfnamefont{S.~K.} \bibnamefont{Donaldson}},
  \bibinfo{journal}{Commun. Math. Phys.} \textbf{\bibinfo{volume}{96}},
  \bibinfo{pages}{387} (\bibinfo{year}{1984}).

\bibitem[{\citenamefont{Hurtubise}(1985)}]{Hurt-1985}
\bibinfo{author}{\bibfnamefont{J.}~\bibnamefont{Hurtubise}},
  \bibinfo{journal}{Commun. Math. Phys.} \textbf{\bibinfo{volume}{100}},
  \bibinfo{pages}{191} (\bibinfo{year}{1985}).

\bibitem[{\citenamefont{Takhtajan}(1992)}]{Takhtajan-1992}
\bibinfo{author}{\bibfnamefont{L.~A.} \bibnamefont{Takhtajan}},
  \bibinfo{journal}{Theor. Math. Phys.} \textbf{\bibinfo{volume}{93}},
  \bibinfo{pages}{1308} (\bibinfo{year}{1992}).

\bibitem[{\citenamefont{Stuart}(1994)}]{Stuart-1994}
\bibinfo{author}{\bibfnamefont{D.}~\bibnamefont{Stuart}},
  \bibinfo{journal}{Commun. Math. Phys.} \textbf{\bibinfo{volume}{166}},
  \bibinfo{pages}{149} (\bibinfo{year}{1994}).

\bibitem[{\citenamefont{Gibbons and Manton}(1986)}]{Manton-1986}
\bibinfo{author}{\bibfnamefont{G.~W.} \bibnamefont{Gibbons}} \bibnamefont{and}
  \bibinfo{author}{\bibfnamefont{N.~S.} \bibnamefont{Manton}},
  \bibinfo{journal}{Nucl. Phys. B} \textbf{\bibinfo{volume}{274}},
  \bibinfo{pages}{183} (\bibinfo{year}{1986}).

\bibitem[{\citenamefont{Schroers}(1991)}]{Schroer-1991}
\bibinfo{author}{\bibfnamefont{B.~J.} \bibnamefont{Schroers}},
  \bibinfo{journal}{Nucl. Phys. B} \textbf{\bibinfo{volume}{367}},
  \bibinfo{pages}{177} (\bibinfo{year}{1991}).

\bibitem[{\citenamefont{Asselmeyer-Maluga and
  Kr\'ol}(2012{\natexlab{c}})}]{AsselmKrol-2012e}
\bibinfo{author}{\bibfnamefont{T.}~\bibnamefont{Asselmeyer-Maluga}}
  \bibnamefont{and} \bibinfo{author}{\bibfnamefont{J.}~\bibnamefont{Kr\'ol}}
  (\bibinfo{year}{2012}{\natexlab{c}}), \bibinfo{note}{arXiv:1201.3787}.

\bibitem[{\citenamefont{Balakin et~al.}(2008)\citenamefont{Balakin, Dehnen, and
  Zayats}}]{Dehnen-2008}
\bibinfo{author}{\bibfnamefont{A.~B.} \bibnamefont{Balakin}},
  \bibinfo{author}{\bibfnamefont{H.}~\bibnamefont{Dehnen}}, \bibnamefont{and}
  \bibinfo{author}{\bibfnamefont{A.~E.} \bibnamefont{Zayats}},
  \bibinfo{journal}{Annals Phys.} \textbf{\bibinfo{volume}{323}},
  \bibinfo{pages}{2183} (\bibinfo{year}{2008}),
  \bibinfo{note}{arXiv:0804.2196}.

\bibitem[{\citenamefont{Castelnovo et~al.}(2008)\citenamefont{Castelnovo,
  Moessner, and Sondhi}}]{Castelnovo-2008}
\bibinfo{author}{\bibfnamefont{C.}~\bibnamefont{Castelnovo}},
  \bibinfo{author}{\bibfnamefont{R.}~\bibnamefont{Moessner}}, \bibnamefont{and}
  \bibinfo{author}{\bibfnamefont{S.~L.} \bibnamefont{Sondhi}},
  \bibinfo{journal}{Nature} \textbf{\bibinfo{volume}{451}}, \bibinfo{pages}{42}
  (\bibinfo{year}{2008}).

\bibitem[{\citenamefont{Jaubert and Holdsworth}(2009)}]{Jaubert-2009}
\bibinfo{author}{\bibfnamefont{L.~D.~C.} \bibnamefont{Jaubert}}
  \bibnamefont{and} \bibinfo{author}{\bibfnamefont{P.~C.~W.}
  \bibnamefont{Holdsworth}}, \bibinfo{journal}{Nature Phys.}
  \textbf{\bibinfo{volume}{5}}, \bibinfo{pages}{258 } (\bibinfo{year}{2009}).

\bibitem[{\citenamefont{Giblin}(2011)}]{Giblin-2011}
\bibinfo{author}{\bibfnamefont{S.~R.} \bibnamefont{Giblin}},
  \bibinfo{journal}{Nature Phys.} \textbf{\bibinfo{volume}{7}},
  \bibinfo{pages}{252 } (\bibinfo{year}{2011}).

\bibitem[{\citenamefont{Kawaguchi et~al.}(2008)\citenamefont{Kawaguchi, Nitta,
  and Ueda}}]{Kawa-2008}
\bibinfo{author}{\bibfnamefont{Y.}~\bibnamefont{Kawaguchi}},
  \bibinfo{author}{\bibfnamefont{M.}~\bibnamefont{Nitta}}, \bibnamefont{and}
  \bibinfo{author}{\bibfnamefont{M.}~\bibnamefont{Ueda}},
  \bibinfo{journal}{Phys. Rev. Lett.} \textbf{\bibinfo{volume}{100}},
  \bibinfo{pages}{180403} (\bibinfo{year}{2008}).

\bibitem[{\citenamefont{Pietil\"a and M\"ott\"onen}(2009)}]{Piet-2009}
\bibinfo{author}{\bibfnamefont{V.}~\bibnamefont{Pietil\"a}} \bibnamefont{and}
  \bibinfo{author}{\bibfnamefont{M.}~\bibnamefont{M\"ott\"onen}},
  \bibinfo{journal}{Phys. Rev. Lett.} \textbf{\bibinfo{volume}{102}},
  \bibinfo{pages}{080403} (\bibinfo{year}{2009}).

\bibitem[{\citenamefont{Savage and Ruostekoski}(2003)}]{Savage-2003}
\bibinfo{author}{\bibfnamefont{C.~M.} \bibnamefont{Savage}} \bibnamefont{and}
  \bibinfo{author}{\bibfnamefont{J.}~\bibnamefont{Ruostekoski}},
  \bibinfo{journal}{Phys. Rev. A} \textbf{\bibinfo{volume}{68}},
  \bibinfo{pages}{043604} (\bibinfo{year}{2003}).

\bibitem[{\citenamefont{Asselmeyer-Maluga and Kr{\'o}l}(2011)}]{AsselmKrol11}
\bibinfo{author}{\bibfnamefont{T.}~\bibnamefont{Asselmeyer-Maluga}}
  \bibnamefont{and} \bibinfo{author}{\bibfnamefont{J.}~\bibnamefont{Kr{\'o}l}}
  (\bibinfo{year}{2011}), \bibinfo{note}{arXiv:1107.0650}.

\bibitem[{\citenamefont{Kr\'ol}(2012)}]{QG-2012}
\bibinfo{author}{\bibfnamefont{J.}~\bibnamefont{Kr\'ol}}, in
  \emph{\bibinfo{booktitle}{Quantum Gravity}} (\bibinfo{publisher}{Rodrigo
  Sobreiro (Ed.), ISBN: 978-953-51-0089-8, InTech, Available from:
  http://www.intechopen.com/articles/show/title/quantum-gravity-insights-from-smooth-4-geometries-on-trivial-r4},
  \bibinfo{year}{2012}).

\end{thebibliography}
\end{document}